\documentclass{aa}

\newcommand{\lta}{\;
  \raise0.3ex\hbox{$<$\kern-0.75em\raise-1.1ex\hbox{$\sim$
  }}\;\hskip-2pt }
\newcommand{\gta}{\;
  \raise0.3ex\hbox{$>$\kern-0.75em\raise-1.1ex\hbox{$\sim$
  }}\;\hskip-2pt }
\newcommand{\apropto}{\;
  \raise0.3eZx\hbox{$\propto$\kern-0.75em\raise-1.1ex\hbox{$\sim$
  }}\;\hskip-2pt }
\usepackage[dvips]{graphicx}
\usepackage{graphicx}
\usepackage{ulem}
\usepackage{txfonts}
\usepackage{natbib}
\bibpunct[; ]{(}{)}{;}{a}{}{}
\begin{document}
\title{Multiscale magnetic fields in spiral galaxies: evolution and reversals}
\author{D. Moss\inst{1}\fnmsep
\thanks{Corresponding author: \email{moss@ma.man.ac.uk}}
R. Stepanov\inst{2}, T.\,G. Arshakian\inst{3,4}, R. Beck\inst{3}, M.
Krause\inst{3}, and D. Sokoloff\inst{5}}
\titlerunning{Magnetic fields in spiral galaxies}
\authorrunning{D. Moss et al.}
\institute{School of Mathematics, University of Manchester, Manchester M13 9PL, UK
\and
Institute of Continuous Media Mechanics, Korolyov
str. 1, 614061 Perm, Russia,
\and
MPI f\"ur Radioastronomie, Auf dem H\"ugel 69, 53121
Bonn, Germany
\and
    Byurakan Astrophysical Observatory, Byurakan 378433,
    Armenia and Isaac Newton Institute of Chile, Armenian
    Branch
\and
Department of Physics, Moscow State University, Russia}
\date{Received ?????; accepted ??????}

% \abstract{}{}{}{}{}
% 5 {} token are mandatory

\abstract
%{Text of context}
{Magnetic fields in nearby, star-forming galaxies reveal large-scale
patterns, as predicted by dynamo models, but also a variety of
small-scale structures. In particular, a large-scale field reversal
may exist in the Milky Way while no such reversals have been
observed so far in external galaxies.}
%{Text of aims}
{The effects of star-forming regions
of galaxies need to be included when modelling the
evolution of their magnetic fields, which can then be compared to
future radio polarization observations. The conditions
leading to large-scale field reversals also need clarification.}
%{Text of methods}
{Our simplified model of field evolution in isolated
disc galaxies includes a standard mean-field dynamo and continuous
injection of turbulent fields (the effect of supernova
explosions) in discrete star forming regions by implicit
small-scale dynamo action. Synthetic maps of radio synchrotron
emission and Faraday rotation measures are computed for galaxies at
different evolutionary stages.}
%{Text of results}
{A large-scale dynamo is essential to obtain regular
large-scale spiral magnetic fields, as observed in many galaxies.
These appear, on kpc scales in near energy equilibrium with the turbulence,
after 1--2\,Gyr (corresponding to redshift about
$4-3$). The injection of turbulent fields generates small-scale
field structures. Strong injected
small-scale fields and a large dynamo number (e.g. rapid rotation)
of a galaxy favour the generation of field reversals. Depending on
the model parameters, large-scale field reversals may persist over
many Gyrs and can survive until the present epoch. Significant
polarized radio synchrotron emission from young galaxies is expected
at redshift $\le 4$. Faraday rotation measures (RM) are
crucial to detect field reversals. Large-scale RM patterns of
rotation measures can be observed at redshift $ \le 3$.}
%{Text of conclusions}
{Our model can explain the general form of axisymmetric spiral
fields with many local distortions, as observed in nearby galaxies.
For a slightly different choice of parameters, large-scale field
reversals can persist over the lifetime of a galaxy. Comparing our
synthetic radio maps with future observations of distant galaxies
with the planned Square Kilometre Array (SKA) and its precursors
will allow testing and refinement of models of magnetic field
evolution.}


\maketitle
%
%________________________________________________________________

\section{Introduction}

Magnetic fields of nearby spiral galaxies have been investigated
intensively since the 1980s. The strength and configuration of their
large-scale components and the scale and strength of the small-scale
magnetic field were addressed observationally, while dynamo theory
provided models for their origin and evolution (see e.g. Beck et al.
1996, Widrow 2002 and Kulsrud \& Zweibel 2008 for reviews). On the
other hand, this view of the problem is constrained by the limited
available sample of nearby spiral galaxies and does not address many
problems which are important from the astrophysical viewpoint, for
example the relation between galactic morphology (spiral arms, bars,
halos, interactions) and the magnetic fields, or the occurrence
of large-scale field reversals. Perhaps even more important is the
problem of the magnetic field evolution in the first galaxies to
form. This is the topic of this paper. 

However, our knowledge
concerning the details of hydrodynamics and magnetic fields in the
earliest galaxies is very limited and substantially constrains the
direct numerical simulations of the magnetic field evolution in
these galaxies. This motivated Arshakian et al. (2009, 2011) to
exploit simple semi-qualitative estimates taken from conventional
galactic dynamo models to derive observational tests for magnetic
fields in evolving galaxies.

The new generation of radio telescopes (under construction or under
development) such as the Square Kilometre Array (SKA) and its
precursors opens a broader perspective, that will enable a
substantial enlargement of the variety of galaxies with known
magnetic field configurations, and will address many points which
have remained inaccessible up to now (Beck 2010). In particular, a
natural idea here is to address theoretically magnetic fields in
galaxies with as large redshifts as possible, in order to understand
the magnetic field evolution in the earliest galaxies with the intention of
observing them with the SKA (e.g. Murphy 2009). The key issue here is
to suggest a set of observational tests for the first galaxies that
lie within the capabilities of this new generation of telescopes.

Numerical simulations using several approaches have illuminated particular
features of the magnetic field evolution during the course of galactic
formation and evolution (e.g. Sur et al. 2007, Gressel et al. 2008,
Hanasz et al. 2009, Wang \& Abel 2009, Mantere et al. 2010,
Schleicher et al. 2010, Kulpa-Dybel et al. 2011). Detailed
models attempt to simulate a wide range of physical processes, and
contain many governing parameters which need to be specified in order
to describe a particular galaxy or a morphological type of galaxies.
The complexity of parametric space and the huge processing times
make an extensive inspection of their parameter space impractical.

The aim of this paper is to introduce a new type of mean-field
galactic dynamo model which solves a relatively detailed set of
mean-field dynamo equations coupled to the continuous injection of
small-scale fields, and improves on the semi-qualitative estimates
by Arshakian et al. (2009). On the other hand, our model is much
simpler than those involving direct numerical simulations. The model
also allows us to mimic to some extent the contribution of
small-scale magnetic fields to the overall field configuration, in
contrast to conventional mean-field dynamo models which take into
account only the large-scale component. The relatively simple nature
of our model makes a fairly detailed investigation of the parameter
space a realistic proposition, and allows us to isolate robust
features of the dynamo-generated magnetic fields which are suitable
for observational identification. Future investigation of these
results by more detailed models are highly welcome.

Our new model is situated between qualitative and semi-analytical
estimates and direct numerical simulations. It is based on what has become
known as the no-$z$ model, suggested by Subramanian \& Mestel (1993)
and Moss (1995, 1997). The idea is to average the magnetic fields in
the galactic disc over various heights $z$ above and below the
central galactic plane. This average becomes dependent on the
galacto-centric radius $r$ and the azimuthal angle $\phi$ but is
independent of  the perpendicular coordinate $z$.
The resulting 2D problem is then numerically
affordable. As a seed field for the model we take a random
seed field, and include continuous generation of
small-scale random magnetic fields at the lowest scale resolvable by
our numerical code (and similar to the seed field in strength and
structure) during the galactic evolution. Then we link
the dynamo and star-formation history to simulate the cosmological
evolution of the magnetic structure.

Our model is based on a 2D set of dynamo equations which yields
a 2D magnetic field configuration at the equatorial plane of a
galaxy. Then we restore the vertical magnetic field component at
this plane from the divergence-free condition and extrapolate the
components from the equatorial plane to the whole disc. As a result
we get a 3D model after explicit solution of 2D equations, which is
computationally efficient.

Finally, from the magnetic field structures we simulate the expected
total and polarized emission, and Faraday rotation measures (RMs) in
these disc galaxies.

The main message obtained from the model can be summarized as
follows. We recognize two phases of the evolution of galactic
magnetic fields. Firstly, a large-scale magnetic field configuration
develops from the random seed field. This stage is quite short and
takes 1--2\,Gyr. Secondly, the magnetic field becomes quite regular
and close to an axisymmetric spiral structure, but affected by
fluctuations. The main issue is that we obtain two different types
of large-scale magnetic field configuration: (1) axisymmetric
structures that have the same field direction over the whole
galactic disc, (2) axisymmetric spiral fields that are restricted to
two or even three separate intervals in radius (rings) with the
field reversing direction between neighbouring rings.

\section{A model for the evolution of magnetic fields in galaxies}

A three-phase model for the evolution of magnetic fields in
isolated (without merger) disc galaxies in the context of the
hierarchical structure formation cosmology was developed by
Arshakian et al. (2009). According to this model, in the first phase
weak fields of order $\sim 10^{-18}$\,G were generated by the
Biermann battery mechanism and/or the Weibel instability in the
first dark matter halos. The second phase was manifested by merging
of halos and thermal virialization of protogalaxies. During this
epoch the small-scale dynamo was able to amplify effectively the
turbulent magnetic field up to the energy level of equilibrium with
turbulent kinetic energy ($\sim 10^{-5}$\,G) in a relatively short
timescale of a few hundreds of million years. The third phase started at
the epoch of disc formation (at a redshift of about 10) that
occurred by dissipation of the protogalactic halo. The magnetic
field preserved in the protogalactic halo served as a seed field in
the disc ($\sim 10^{-7}$\,G), which was further amplified by the
mean-field dynamo to the equipartition level ($\sim 10^{-5}$\,G) in
few Gyrs and ordered on scales of up to galactic scale in about
13\,Gyr, this stage lasting until the present epoch. In
semi-analytical models of the evolution of regular magnetic fields
Arshakian et al. (2011) proposed that the configuration of the
initial regular field in the third phase was ``spotty'' over the
galactic disc. The differential rotation initially stretches the
magnetic spots in the azimuthal direction as they appear, and field
reversals can be formed in few Gyrs by the effects of the
large-scale dynamo action, and eventually an axisymmetric field
similar to that of many present-day galaxies.

In this section, we adopt some of the ideas of the
semi-analytical model of Arshakian et al. (2009, 2011) and develop
an evolutionary model of a disc galaxy (phase 3), based on explicit
solutions of the dynamo equations coupled to the star formation
rate.

\subsection{Star-formation and gas turbulence in the disc}
\label{ssec:sfh}

The variations of the star-formation rate (SFR) and the
physical/geometrical parameters of the disc drive the evolution of
regular magnetic fields in galaxies (Arshakian et al. 2011). Star
formation can be triggered by different processes including, for
example, gravitational instability, and the interaction of clouds
and tidal forces in isolated and merging galaxies (Kennicutt et al.
1987; Combes 2005). The rate of SN explosions is proportional to the
SFR. Supernova remnants (SNRs) drive the turbulence of the
interstellar medium (ISM) and determine the characteristic velocity
dispersion of the gas, characterized by the turbulent velocity $v$.
The latter is known to be correlated with SFR for nearby galaxies
(Dib et al. 1996); it is almost constant for low SFRs (up to SFR
values typical of the Milky Way) with $v\approx 10$\,km s$^{-1}$
and grows exponentially at higher SFRs as typical for starburst
galaxies. Star formation in isolated, young galaxies is more
efficient because more gas is available at high redshifts
(redshift $\sim 1.3$, Ryan et al. 2007). SFR is proportional to the
gas mass density (the Kennicutt-Schmidt law) and, hence, the SFR
history of disc galaxies is different for galaxies with different
Hubble types. Modelling of SFR indicates that a constant SFR is
appropriate for late-type spirals (Sd) while the SFR decreases with
galaxy age for earlier types (Sc, Sb, and Sa; Sandage 1986, Kotulla
et al. 2009). For simplicity, we consider an evolving Milky Way-type
galaxy with constant SFR of 1 $M_{\sun}\,{\mathrm{yr}^{-1}}$ and
constant turbulent velocity, $v=10$\,km s$^{-1}$ throughout its
evolution.

Star formation generates supernova explosions which are the main
source of turbulence in the galaxy disc and in turn inject
small-scale magnetic fields. In our model the field injection occurs
at regular time intervals simultaneously at a number of random
locations with a volume filling factor of about 1\%.

\subsection{The dynamo model}
\label{ssec:dynamo_model}

We use a thin disc galaxy code with the "no-$z$" formulation (e.g.
Subramanian \& Mestel 1993; Moss 1995 and subsequent papers), taking
the $\alpha\omega$ approximation. This code solves explicitly for
the field components parallel to the disc plane with the implicit
understanding that the component perpendicular to this plane
(i.e. in the $z$-direction) is
given by the condition $\nabla\cdot {\bf B}=0$, and that the field
has even (quadrupole-like) parity with respect to the disc plane.
The field components parallel to the plane can be considered as
mid-plane values, or as a form of vertical average through the disc
(see, e.g., Moss 1995). The key parameters are the aspect ratio
$\lambda=h/R$, where $h$ corresponds to the semi-thickness of the
warm gas disc and $R$ is its radius, and the dynamo numbers
$R_\alpha=\alpha_0 R/\eta, R_\omega=\Omega_0 R^2/\eta$. $\lambda$
must be a small parameter. $\eta$ is the turbulent
diffusivity, assumed uniform, and $\alpha_0, \Omega_0$ are typical
values of  the $\alpha$-coefficient and angular velocity
respectively. Thus the dynamo equations become  in cylindrical polar coordinates $(r, \phi, z)$

\begin{eqnarray}
\nonumber
\frac{\partial B_r}{\partial t}&=& -R_\alpha B_\phi-\frac{\pi^2}{4} B_r\\&&+\lambda^2\left(\frac{\partial}{\partial r}\left[\frac{1}{r}\frac{\partial}{\partial r}(rB_r)\right]+\frac{1}{r^2}\frac{\partial^2B_r}{\partial\phi^2}-\frac{2}{r^2}\frac{\partial B_\phi}{\partial\phi}\right) ,
\label{evolBr}
\end{eqnarray}
\begin{eqnarray}
\nonumber
\frac{\partial B_\phi}{\partial t}&=& R_\omega r B_r\frac{d\Omega}{dr}-R_\omega\Omega\frac{\partial B_\phi}{\partial \phi}-\frac{\pi^2}{4} B_\phi\\&&+\lambda^2\left(\frac{\partial}{\partial r}\left[\frac{1}{r}\frac{\partial}{\partial r}(rB_\phi)\right] +\frac{1}{r^2}\frac{B_\phi^2}{\partial \phi^2}-\frac{2}{r^2}\frac{\partial B_r}{\partial \phi}\right),
\label{evolBphi}
\end{eqnarray}
 where $z$ does not appear explicitly.
This equation has been calibrated by introduction of the factors
$\pi^2/4$ in the vertical diffusion terms. Of course, in principle
in the $\alpha\omega$  approximation the parameters $R_\alpha,
R_\omega$ can be combined into a single dynamo number $D=R_\alpha
R_\omega$, but for reasons of convenience and clarity of
interpretation we choose to keep them separate.

In the current implementation the code is now written in cartesian
coordinates $x, y, z$, with $z$-axis parallel to the rotation axis.
Length, time and magnetic field are non-dimensionalized in units of
$R$, $h^2/\eta$ and the equipartition field strength $B_{\rm eq}$
respectively.

A naive algebraic $\alpha$-quenching nonlinearity is assumed,
$\alpha=\alpha_0/(1+B^2/B_{\rm eq}^2)$, where $B_{\rm eq}$ is the
strength of the equipartition field in the general disc environment,
not necessarily that in the ``spots'' -- see below. We appreciate
that more sophisticated approaches to the saturation process exist,
and that careful consideration of helicity transport processes is
required to demonstrate that limitation of the large-scale field at
very low levels ("catastrophic quenching") does not occur -- see
e.g. Vishniac \& Cho (2001), Kleeorin et al. (2002, 2003), Sur et
al. (2007). However it does appear that in some cases at least, such
a naive algebraic quenching can reproduce reasonably well the
results from a  more sophisticated treatment (e.g. Kleeorin et al.
2002). Of course, more sophisticated forms of algebraic
alpha-quenching are also possible, such as forms that are non-local
in space or time. We choose to use the simplest possible approach.

In order to implement boundary conditions that ${\bf B }\rightarrow
{\bf 0}$ at the disc boundary (as used in earlier versions of the
code written in polar coordinates), the disc region $x^2+y^2\le 1$
was embedded in a larger computational region $|x|\le x_{\rm b},
|y|\le y_{\rm b}$, where $x_{\rm b}=y_{\rm b}\approx 1.35$. In the
region $x^2+y^2> 1$ outside of the disc the field satisfies the
diffusion equation without dynamo terms. This was found to give
solutions that were small at the disc boundary, and rapidly became
negligible outside of the disc. The standard implementation used a
uniform grid of $229\times 229$ points over $-1 \le x,y \le +1$,
with appropriate additional points in the surrounding "buffer zone".
With a nominal galaxy radius of 10\,kpc, this gives a resolution of
about 88 pc. With $R=10$\,kpc, the assumed disc semi-thickness of
$h=500$\,pc gives $\lambda=0.05$, and the unit of time is then
approximately 0.78\,Gyr. Subsequently, all fields are in units
of the equipartition field $B_{\rm eq}$ (i.e. energy equipartition
with turbulence), unless explicitly stated.

The coefficient $\alpha_0$ is assumed to be uniform in the work
described in this paper. We performed tests with a radially
varying coefficient $\alpha_0=\alpha_0(r)$, but this did not change
our results significantly.

Taking typical galactic values, we can estimate $R_\alpha=O(1)$,
$R_\omega=O(10)$ (i.e. $D\approx 10$). However, a considerable
degree of uncertainty is attached to the conventional estimate
$\eta=10^{26}$ cm$^2$\,s$^{-1}$, and correspondingly to $R_\alpha$
and $R_\omega$.

>From experiments with an arbitrary seed field, the marginal values
are $R_\alpha\approx 0.6$ with $R_\omega=8$, so the critical
dynamo number for this no-$z$ model is $D_{\rm cr} \approx 5$. This
value can be compared with the critical dynamo number $D_{\rm cr}
\approx 7$ which comes from the local disc dynamo problem (Ruzmaikin
et al. 1988) and illustrates the degree of consistency between the
local and no-$z$ models. In any case, the above estimates of
$R_\alpha$ and $R_\omega$ correspond to a somewhat supercritical
dynamo.

Of course these estimates apply to conditions in contemporary spiral
galaxies and may need revision for very young objects -- we will
return to this later in Sect.~\ref{sec:discussion}. Note again that
we use $z$ as a cartesian coordinate, and not as redshift (except in
one instance, where explicitly stated.)

\subsection{The field injection algorithm}
\label{inject}

In our attempt to describe the generation of small-scale fields
in star-forming regions, $n_{\rm sp}$ spot centres $(x_{\rm sp},
y_{\rm sp})$ are generated at random positions within the disc
($r\le 1$). These rotate with the local angular velocity. At time
intervals $dt_{\rm inj}$ a
random field $B_{\rm x,sp}, B_{\rm y,sp}$ is assigned to each of
these positions $(x_{\rm sp}, y_{\rm sp})$. We, slightly arbitrarily,
take $dt_{\rm inj}$ to be about 10 million years. The neighbouring points,
out to a radius $3\,r_{\rm sp}$, are assigned non-zero magnetic
field by assuming a Gaussian distribution of components $B_{\rm x},
B_{\rm y}$ with half-width $r_{\rm sp}$ and central values $B_{\rm
x,sp}, B_{\rm y,sp}$. These spots are assumed to live for a time
$dt_{\rm sp}>> dt_{\rm inj}$. After this time, the old spots
disappear, and a new set of $n_{\rm sp}$ spots is generated using
the above algorithm. The central field strength in each spot,
$(B_{\rm x,sp}^2+B_{\rm y,sp}^2)^{-1}$, is taken from a Gaussian
distribution with central value $B_{\rm inj}$. Note that after
injection, these fields undergo evolution and take part in the
large-scale dynamo action: thus they are a sort of continually
renewed seed field. Thus the process is not equivalent to evolving a
conventional mean-field dynamo, giving a smooth, large-scale field,
and then adding a random field component. For simplicity, all spots
die, and are born, simultaneously.

\begin{figure}
\includegraphics[width=0.45\textwidth]{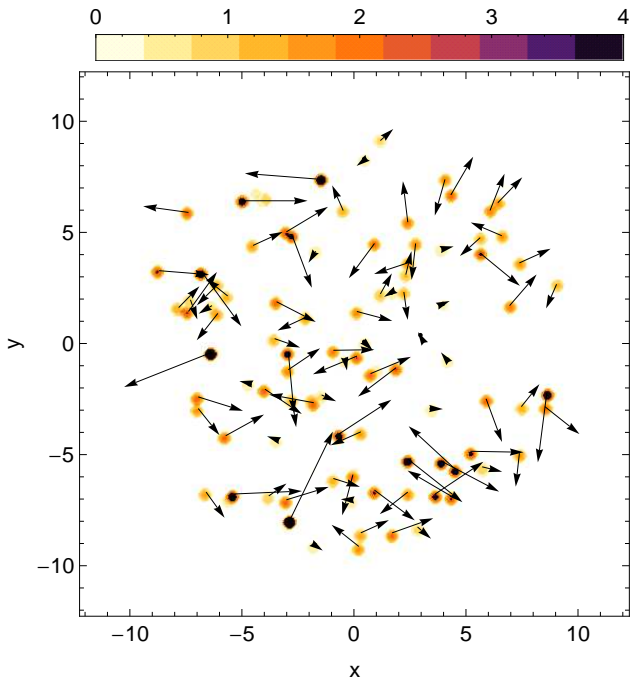}
\caption{ A typical configuration of the initial magnetic field;
only the ${\vec B}$ vectors at the centre of the spots are shown. }
\label{initfield}
\end{figure}

A typical configuration of the initial field is shown in
Fig.~\ref{initfield}. For clarity, this figure plots {\vec B}
vectors at only the central point of each spot. Note that we
consider here magnetic field generation from a small-scale seed
field and use a small-scale initial field that is random,
and distributed in discrete patches. This seed is
supplemented continually by the injection process. We
stress that in many numerical simulations,
a small-scale \cite[e.g.][]{hanasz09} or large-scale seed field,
of dipolar or quadrupolar form
is assumed. Such an initial condition makes implicit
assumptions about the history of the field present as
the galaxy forms. We take a somewhat different viewpoint.

\subsection{Model assumptions}
\label{assumpt}

The initial seed field is assigned by a run of the field
injection algorithm, as described in Sect.~\ref{inject}, i.e. there
is a random, spatially intermittent, initial field. The strength of
this initial field (and also that of the fields subsequently
injected) is determined by the parameter $B_{\rm inj}$ -- see
Sect.~\ref{inject}. Note that because we assume these fields to
result from small-scale dynamo action in the spots ($\sim$
star-forming regions), their strength is near equipartition i.e.
$B_{\rm inj}=O(1)$. In this preliminary study, there is no spatial
weighting of the spot distribution process (Sect.~\ref{inject}), or
$B_{\rm inj}$, nor any time dependence of $B_{\rm inj}$ or $n_{\rm
sp}$. These assumptions correspond to a star formation history that
is spatially and temporally uniform. Clearly a time and/or space
dependence of the injection process could be introduced in further work.

The Gaussian half-width of the spots $r_{\rm sp}=0.01$ (=$100$\,pc
for R=$10$\,kpc) gives an estimate for their filling factor $n_{\rm
sp}r_{\rm sp}$ = $1\%$ for our standard model with $n_{\rm sp}=100$.

\begin{figure}
\includegraphics[width=0.45\textwidth]{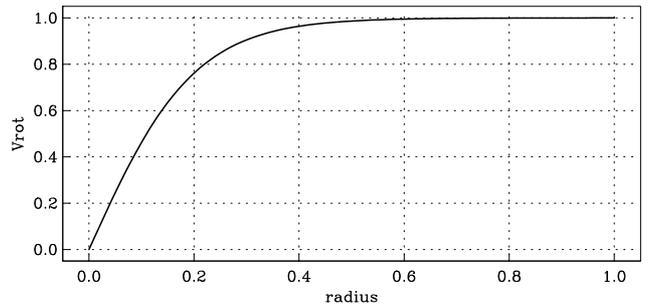}
\caption{The dependence of the scaled rotation velocity on radius for
turnover radius $r_0=0.2$. We choose parameters so that the
flat part of the rotation curve has $V_{\rm rot}\approx 200$ km s$^{-1}$,
and $r=1$ corresponds to the dimensional radius 10\,kpc.}
\label{rotcurve}
\end{figure}

Throughout the paper we take a disc aspect ratio $\lambda=0.05$, and with a
radius of $10$\,kpc the disc semi-thickness is $500$\,pc. For the
rotation curve we take
\begin{equation}
r\frac{d\Omega}{dr}=\Omega_0\left(-\frac{1}{r R_{\rm gal}}\tanh(\frac{r R_{\rm gal}}{r_0})+\frac{1}{r_0\cosh^2(r R_{\rm gal}/r_0)}\right),
\label{rot}
\end{equation}
where $r_0$ corresponds to the turnover radius for the rotational
velocity. We take $r_0=0.2$ and show the dependence of azimuthal
velocity on radius in Fig.~\ref{rotcurve}.

\section{The evolution of magnetic field in selected cases}
\label{evolfield}

We take initial conditions as described above and evolve
Eqs.~(\ref{evolBr}) and (\ref{evolBphi}) until the time $t=17$,
corresponding to a galaxy age $T=13.2$\,Gyr since the epoch at redshift 10.
We take this time $t=17$ as corresponding to the present day epoch. We made
numerous runs with varying parameters.
In order to constrain the number of free parameters, we
keep $R_\alpha=1$ for most of our investigations, and
discuss below the effects of changing $R_\omega$,  $B_{\rm inj}$ and
$n_{\rm sp}$. We also briefly mention at the end of
Sect.~\ref{changeromega} some effects of changes in
$R_\alpha$.

\subsection{The effects of varying the dynamo numbers}
\label{changeromega}

\begin{figure*}
\begin{center}
\begin{tabular}{ll}
(a)
\includegraphics[width=0.41\textwidth]{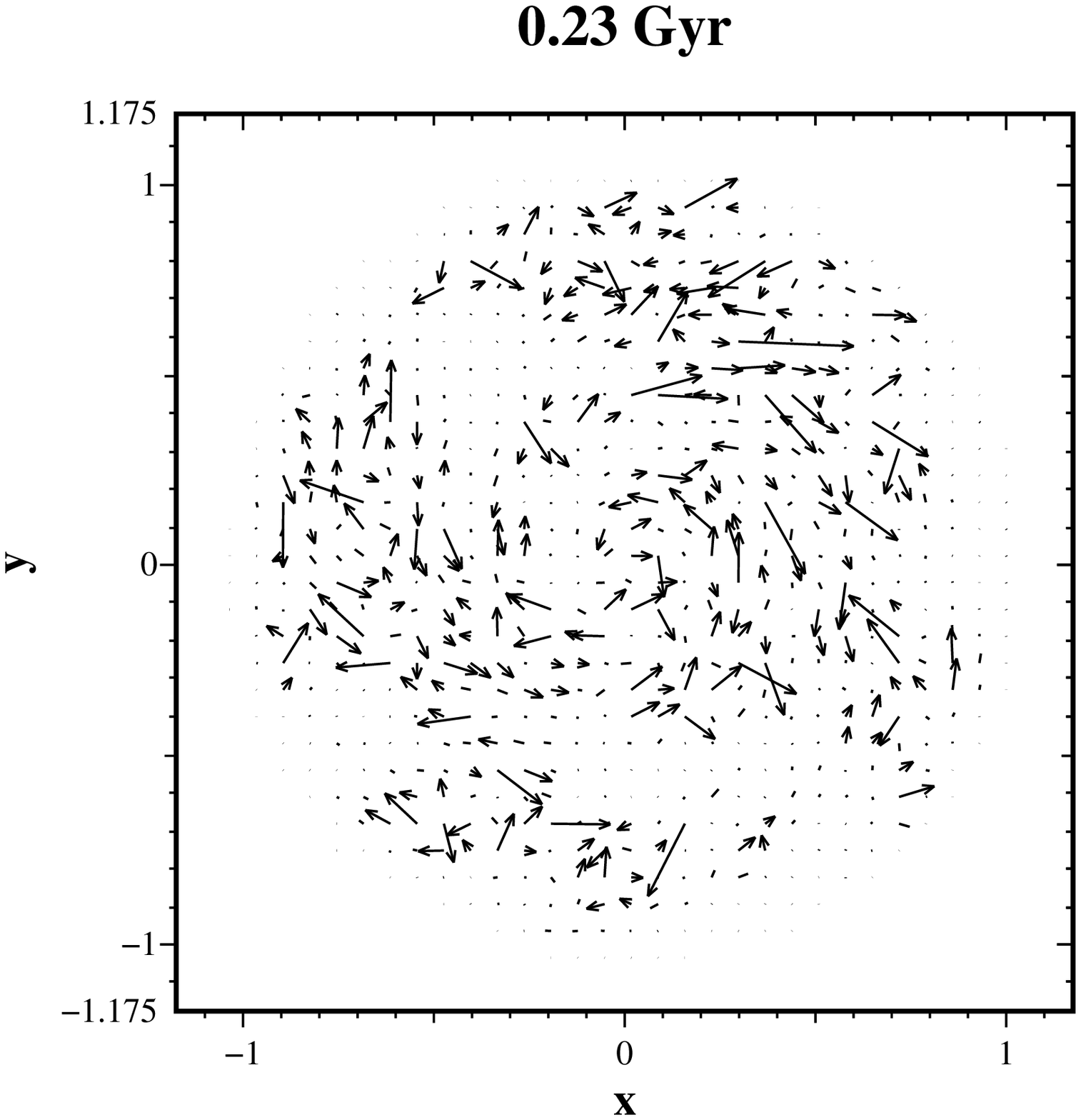}&
\includegraphics[width=0.41\textwidth]{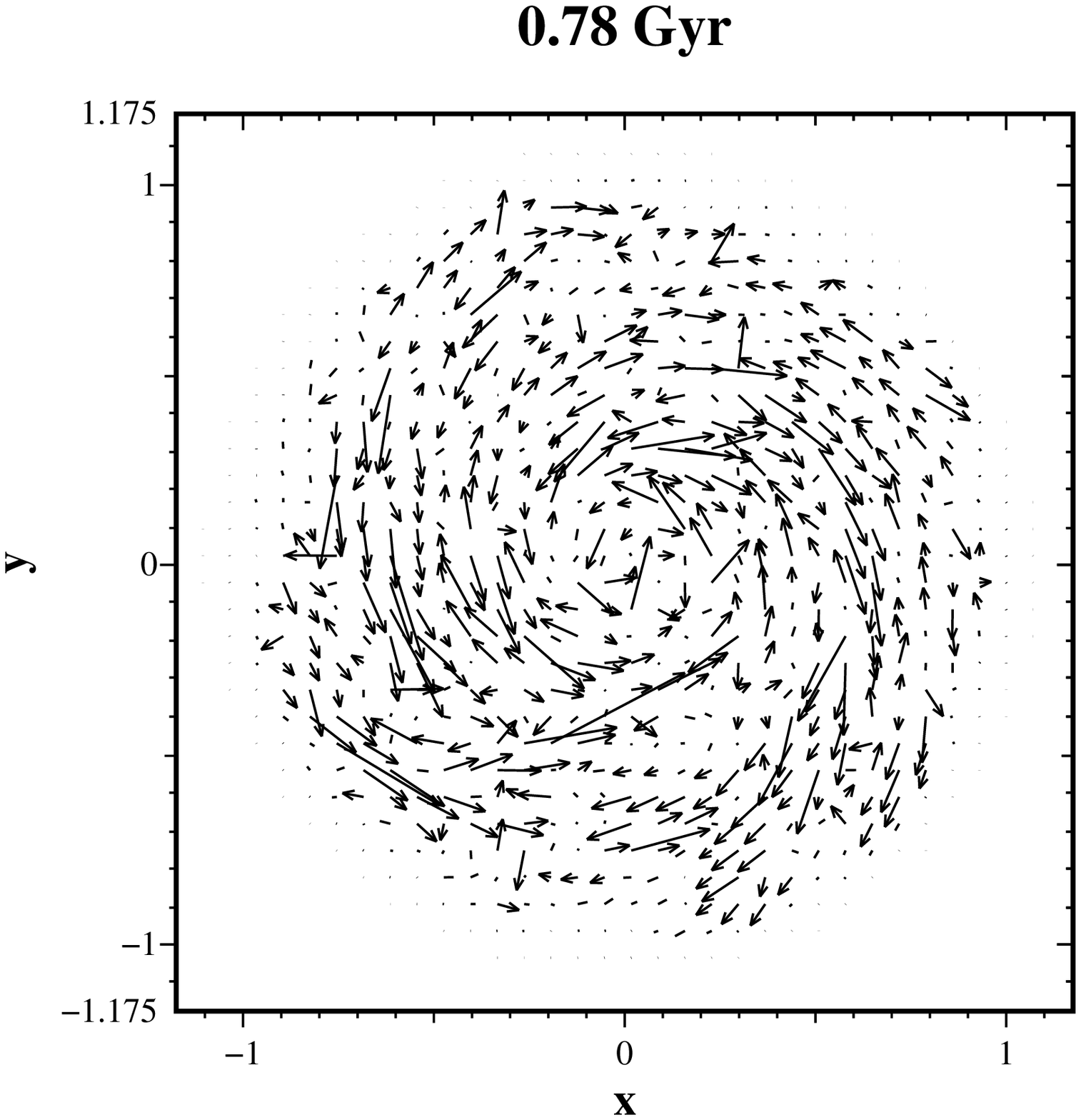}
\\
(b) \includegraphics[width=0.41\textwidth]{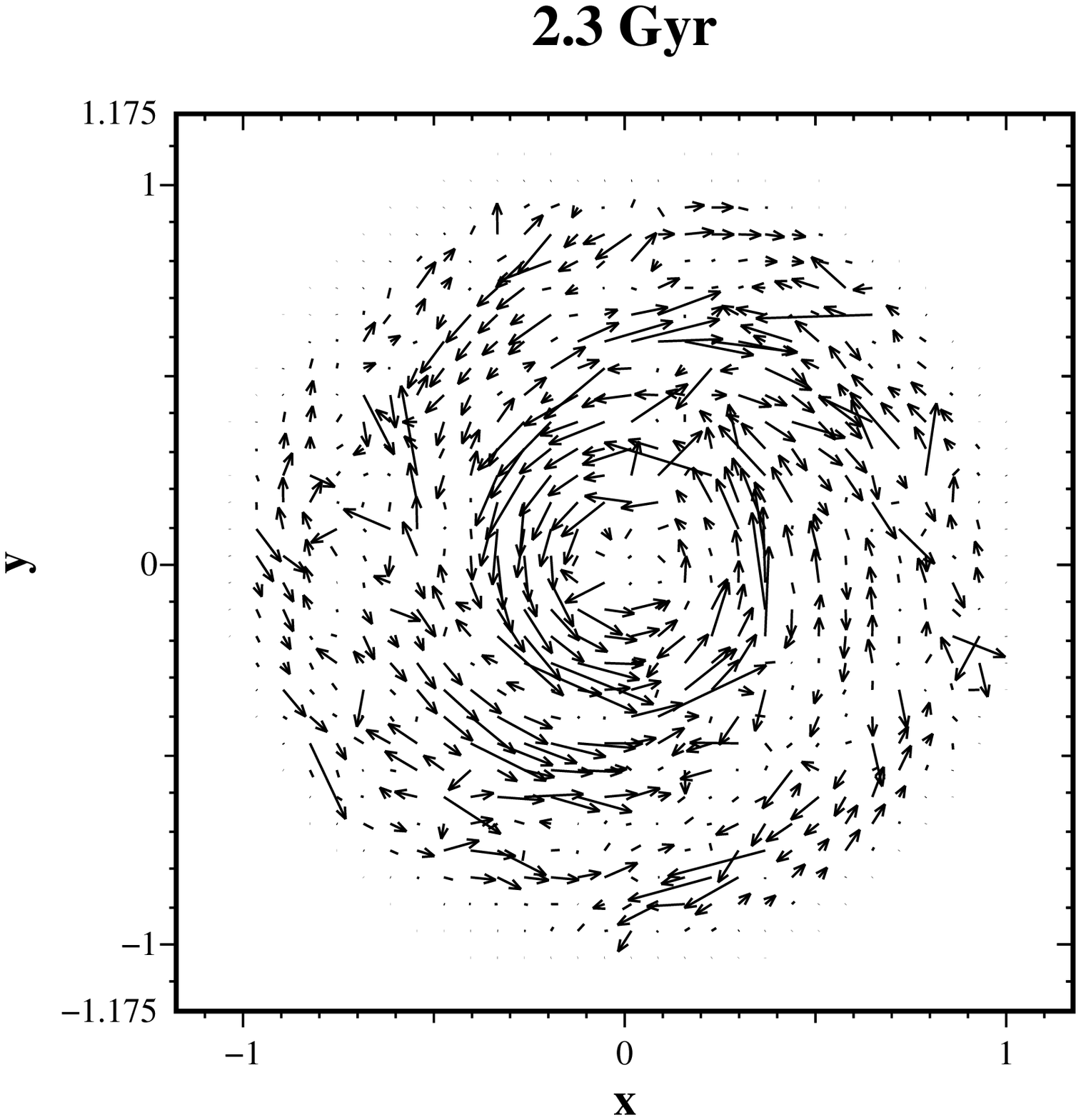} &
\includegraphics[width=0.41\textwidth]{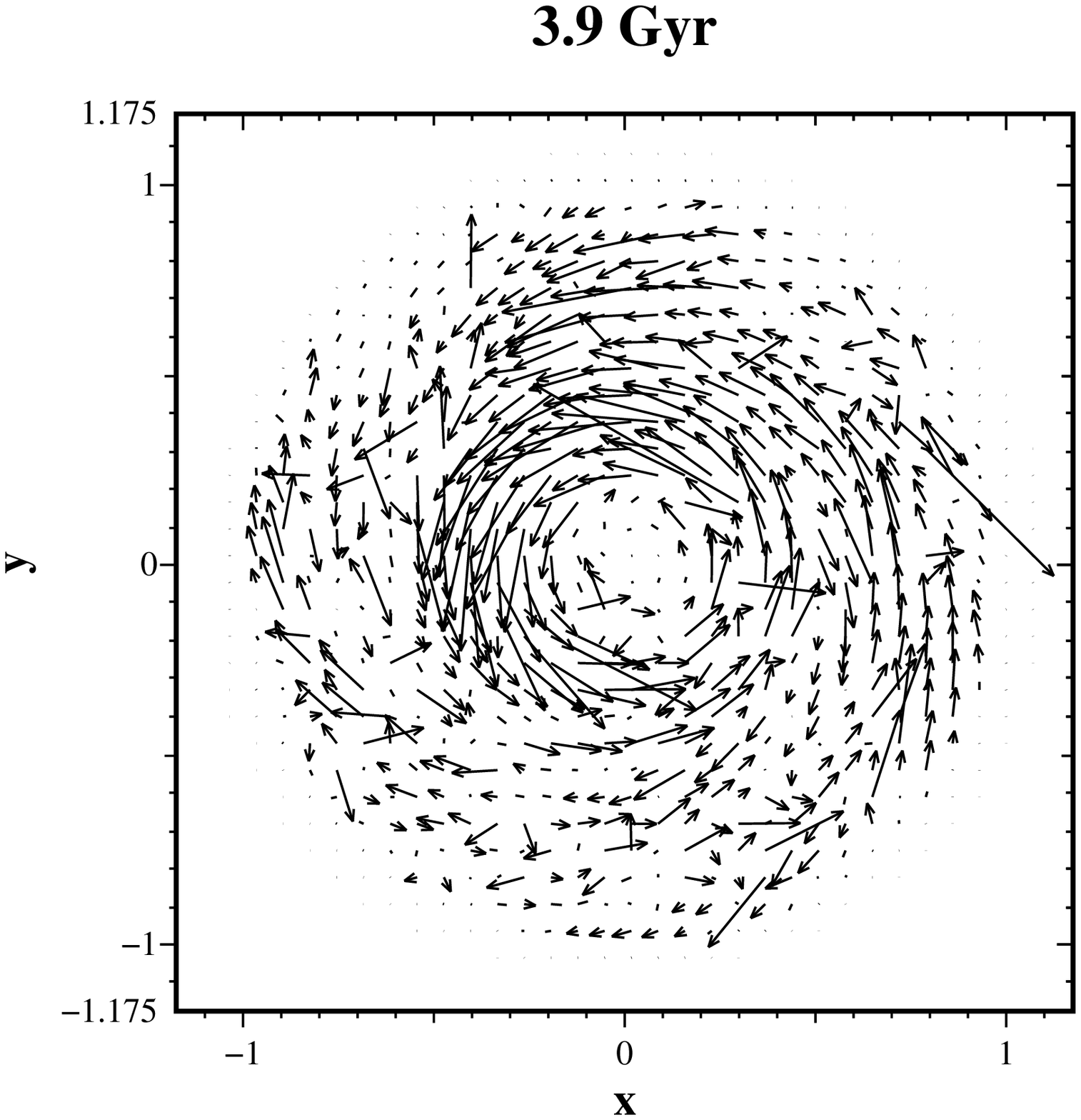}\\
(c)  \includegraphics[width=0.41\textwidth]{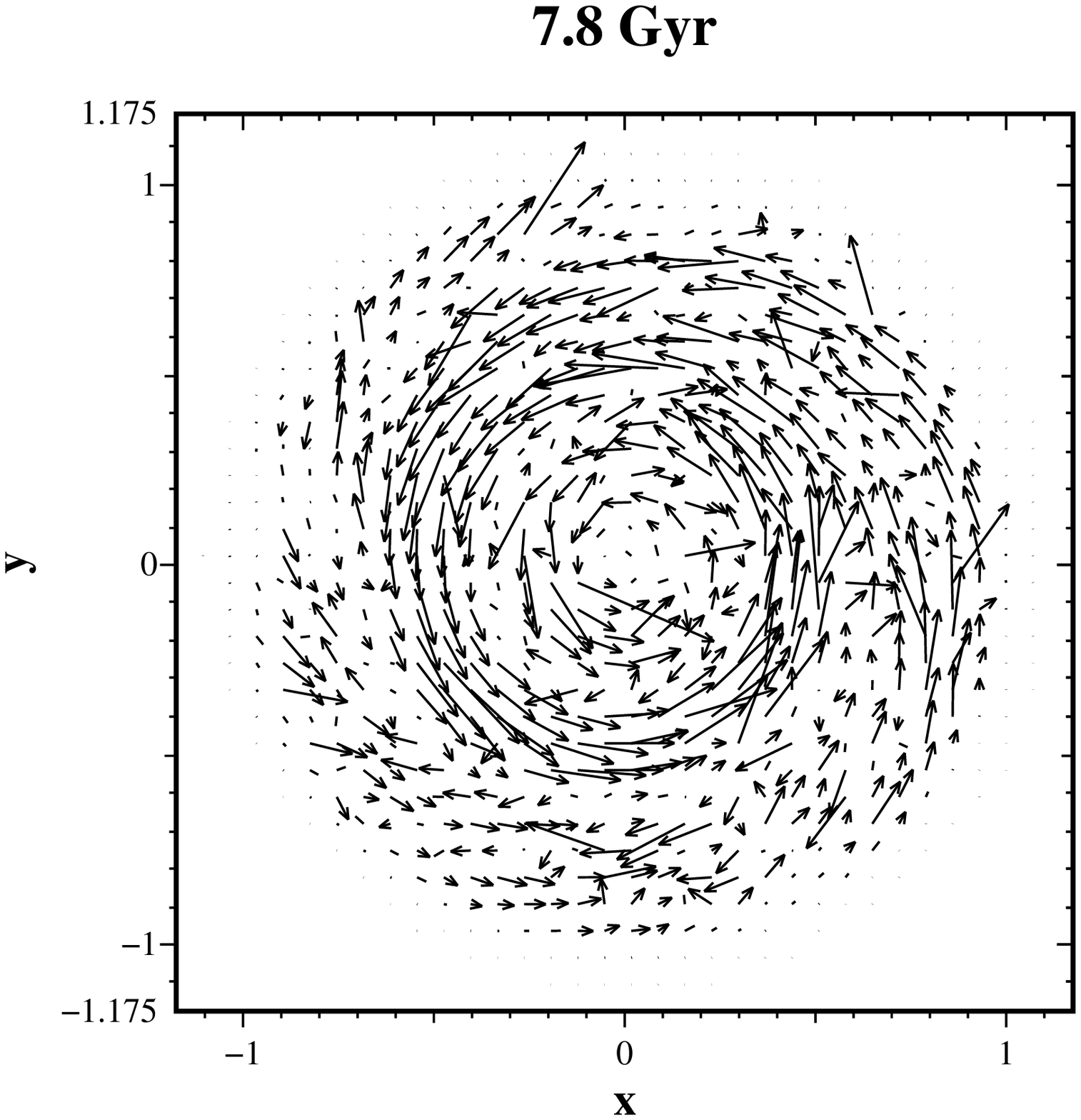} &
  \includegraphics[width=0.41\textwidth]{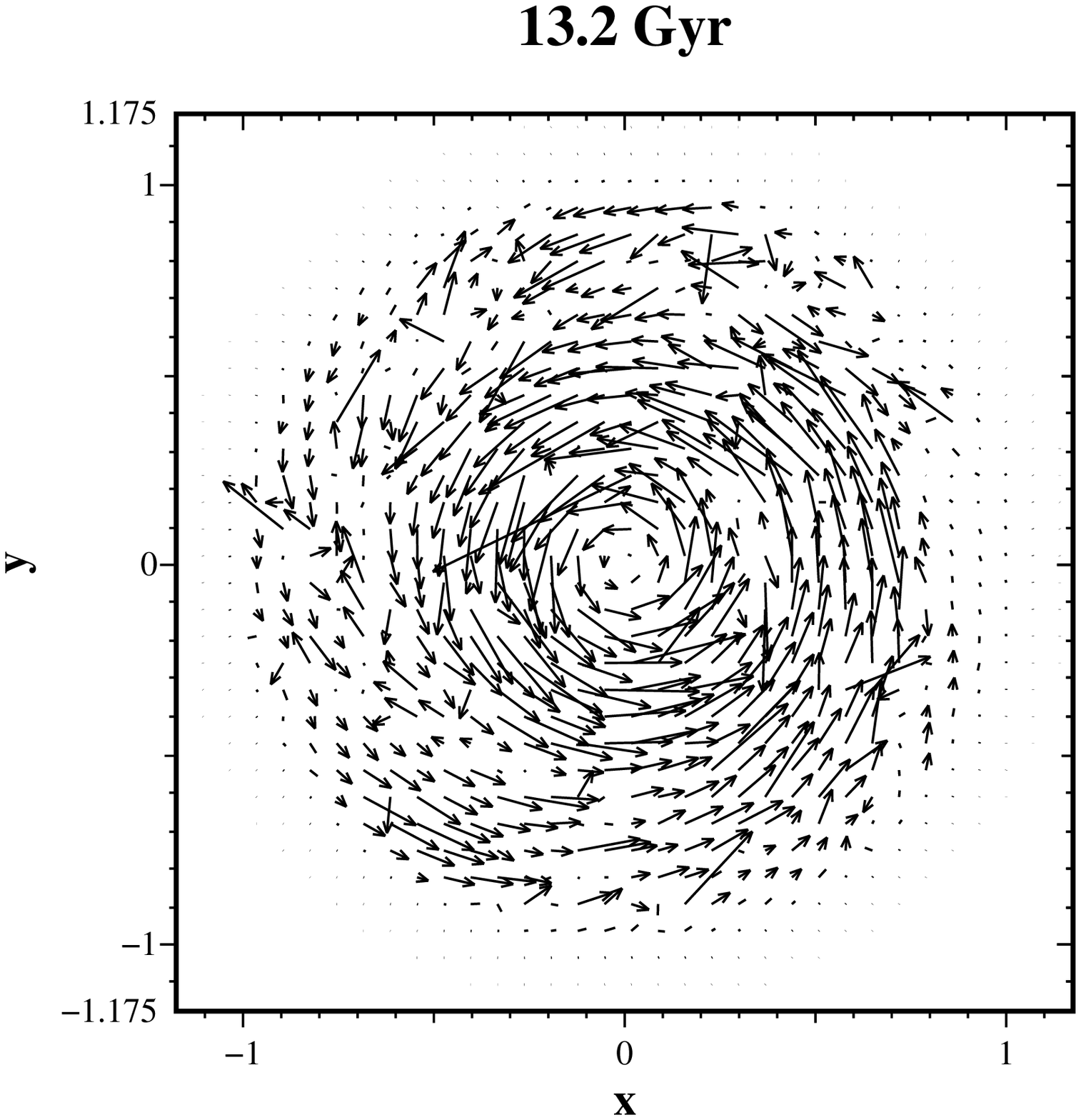}
\end{tabular}
\end{center}
\caption{Field vectors at galaxy ages $T=0.23, 0.78$\,Gyr
(row (a)), $T=2.3, 3.9$\,Gyr (row (b)), and $T=7.8, 13.2$\,Gyr (row
(c)), for the model with $R_\omega=10, R_\alpha=1, B_{\rm inj}=1$.
The vectors give the magnetic field direction, their lengths are
proportional to the magnetic field strengths.
There are no large-scale reversals of field direction visible at the
present day ($t=17$). In this and other plots, the axes are labeled
in units of fractional galactic radius. (Model 138)}
\label{fig:model138}
\end{figure*}

\begin{figure*}
\begin{center}
\begin{tabular}{ll}
(a)  \includegraphics[width=0.41\textwidth]{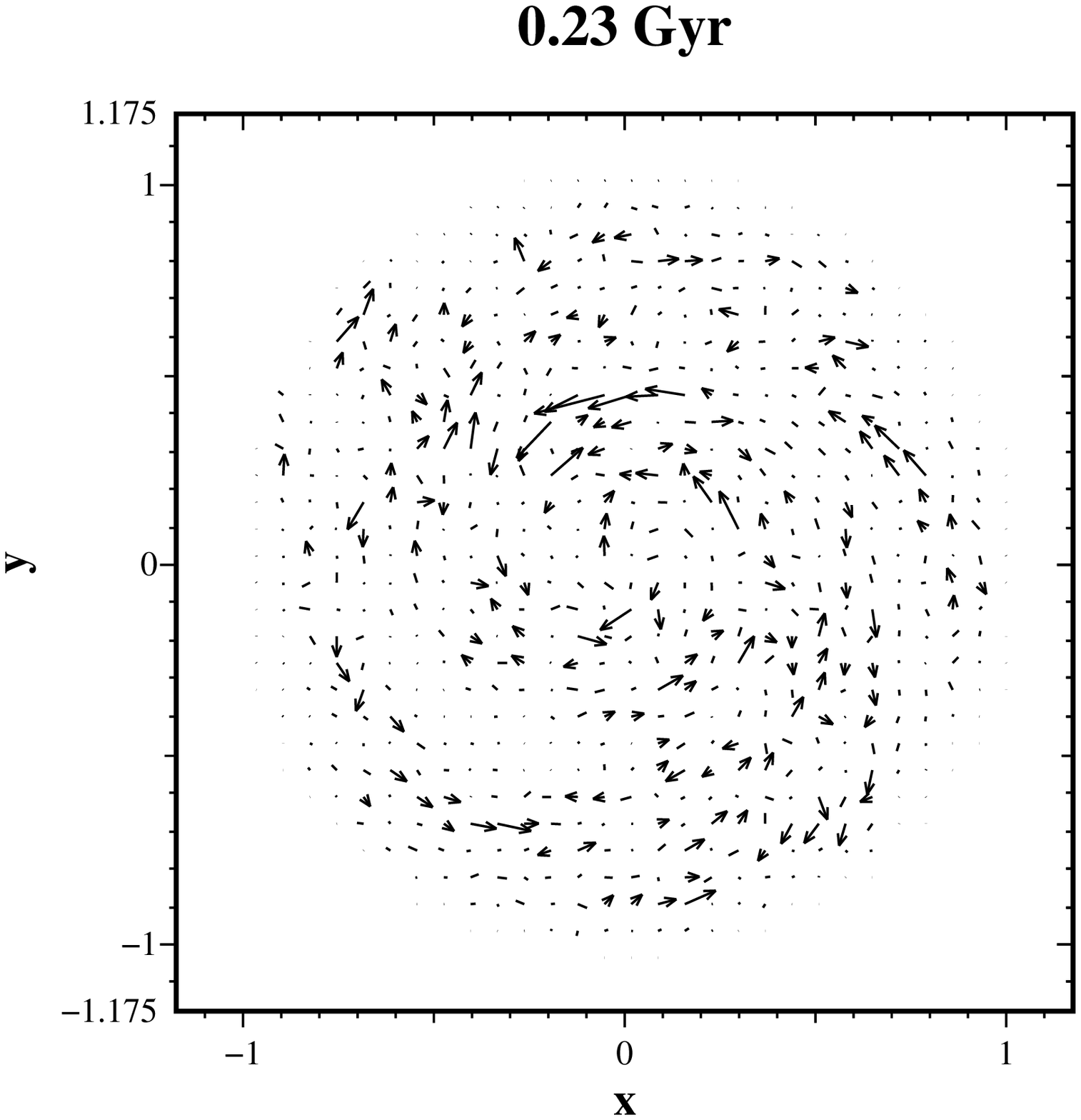} &
\includegraphics[width=0.41\textwidth]{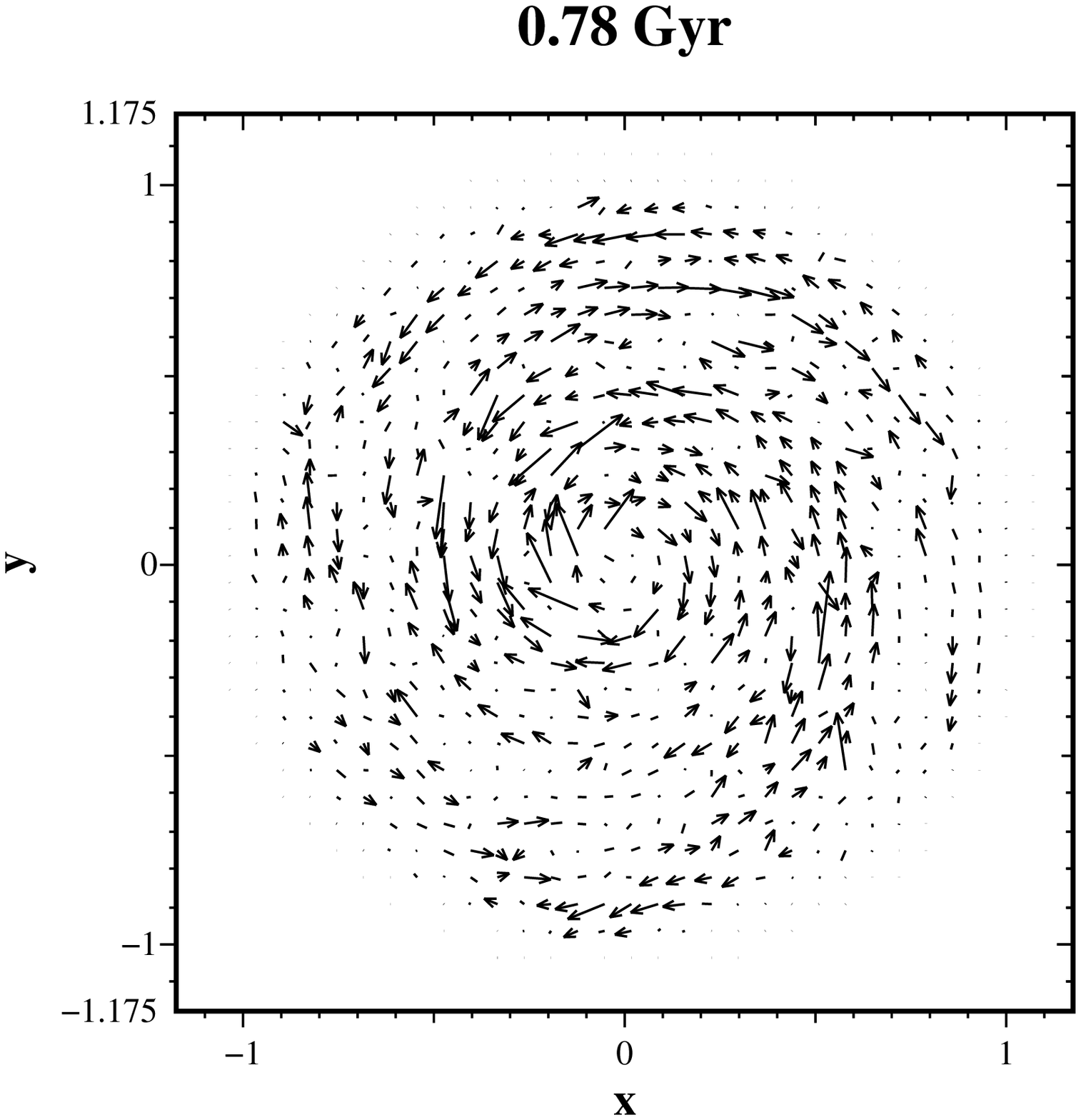}
\\
(b) \includegraphics[width=0.41\textwidth]{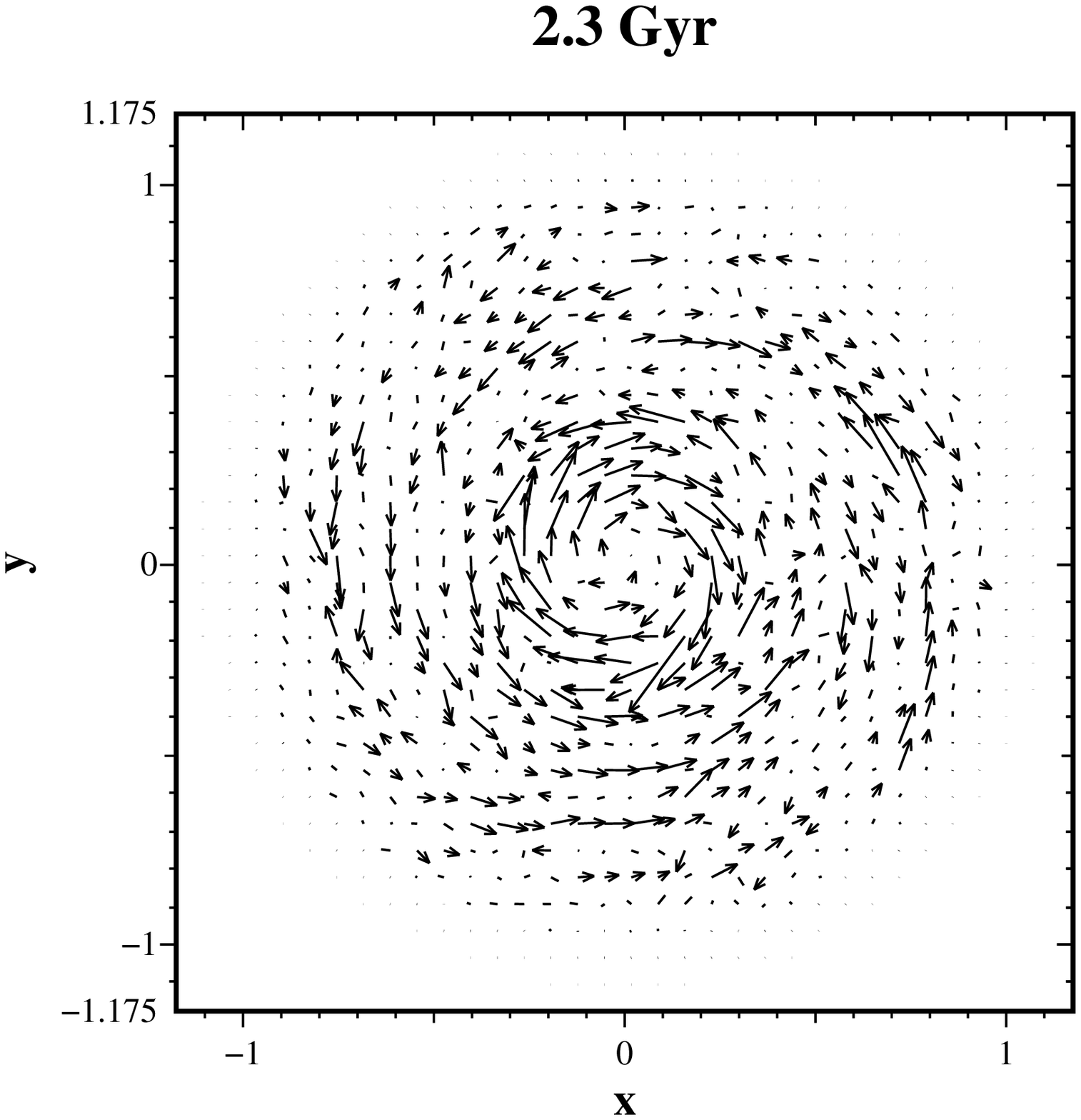} &
\includegraphics[width=0.41\textwidth]{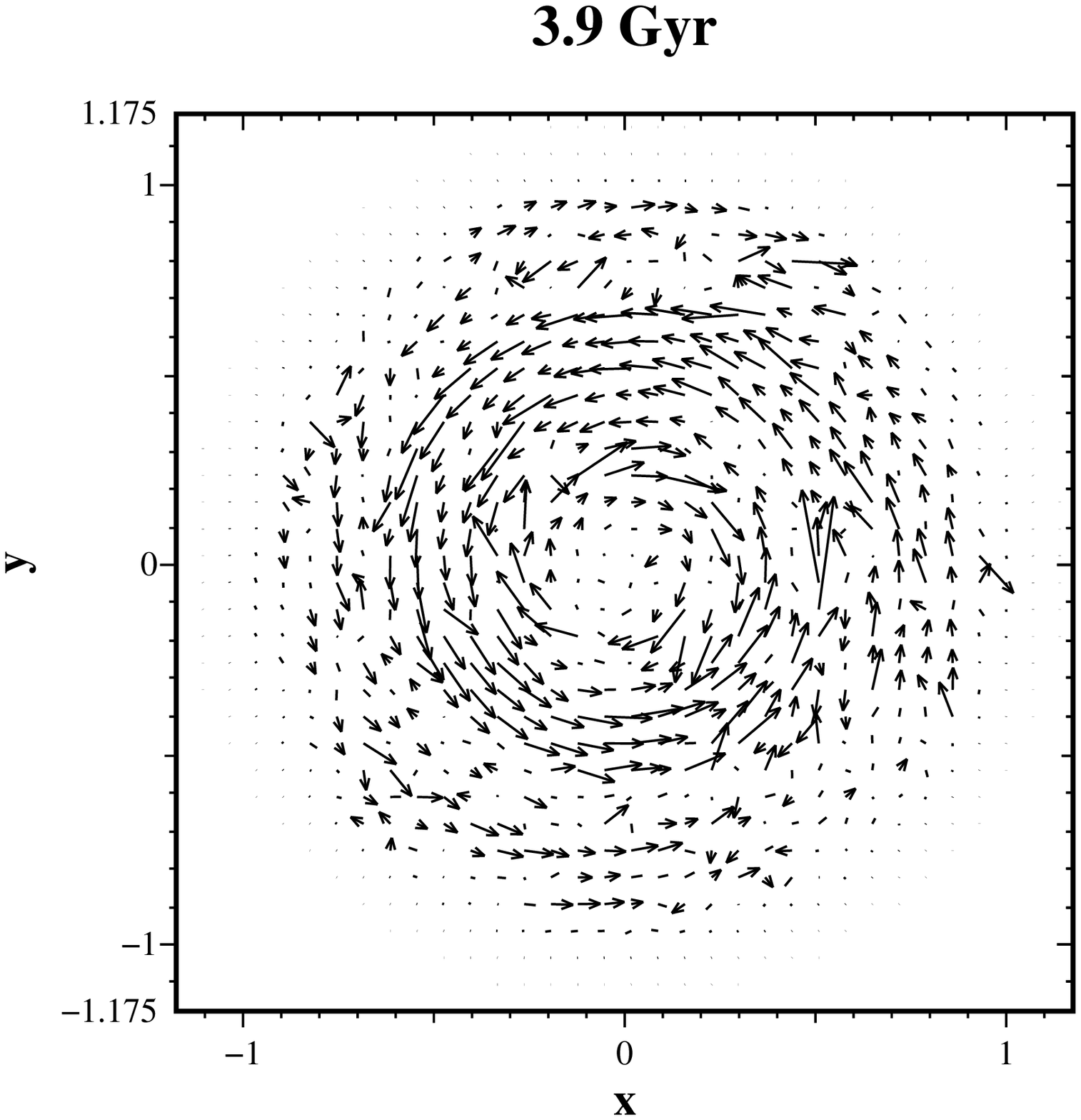}\\
(c)  \includegraphics[width=0.41\textwidth]{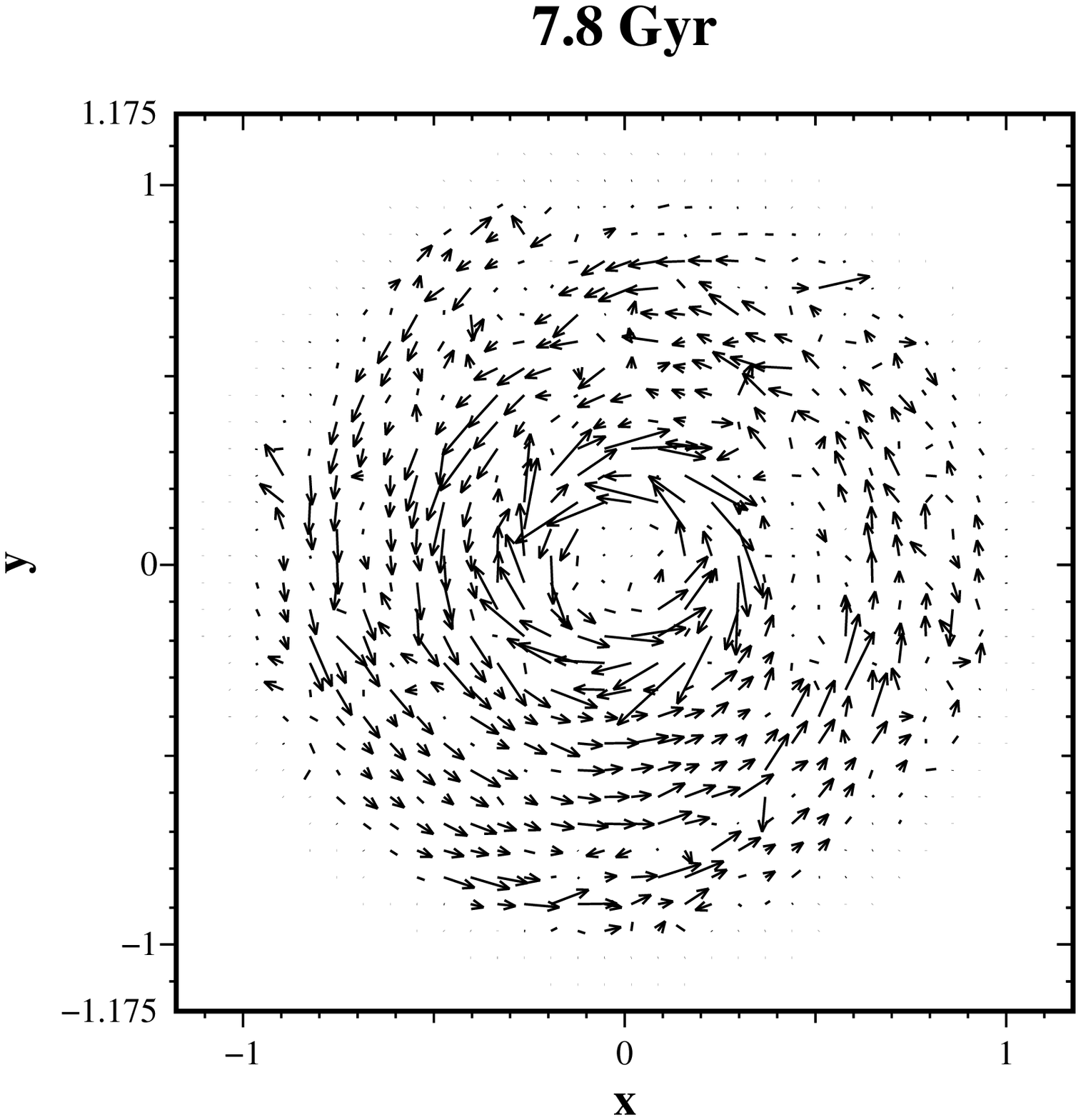} &
\includegraphics[width=0.41\textwidth]{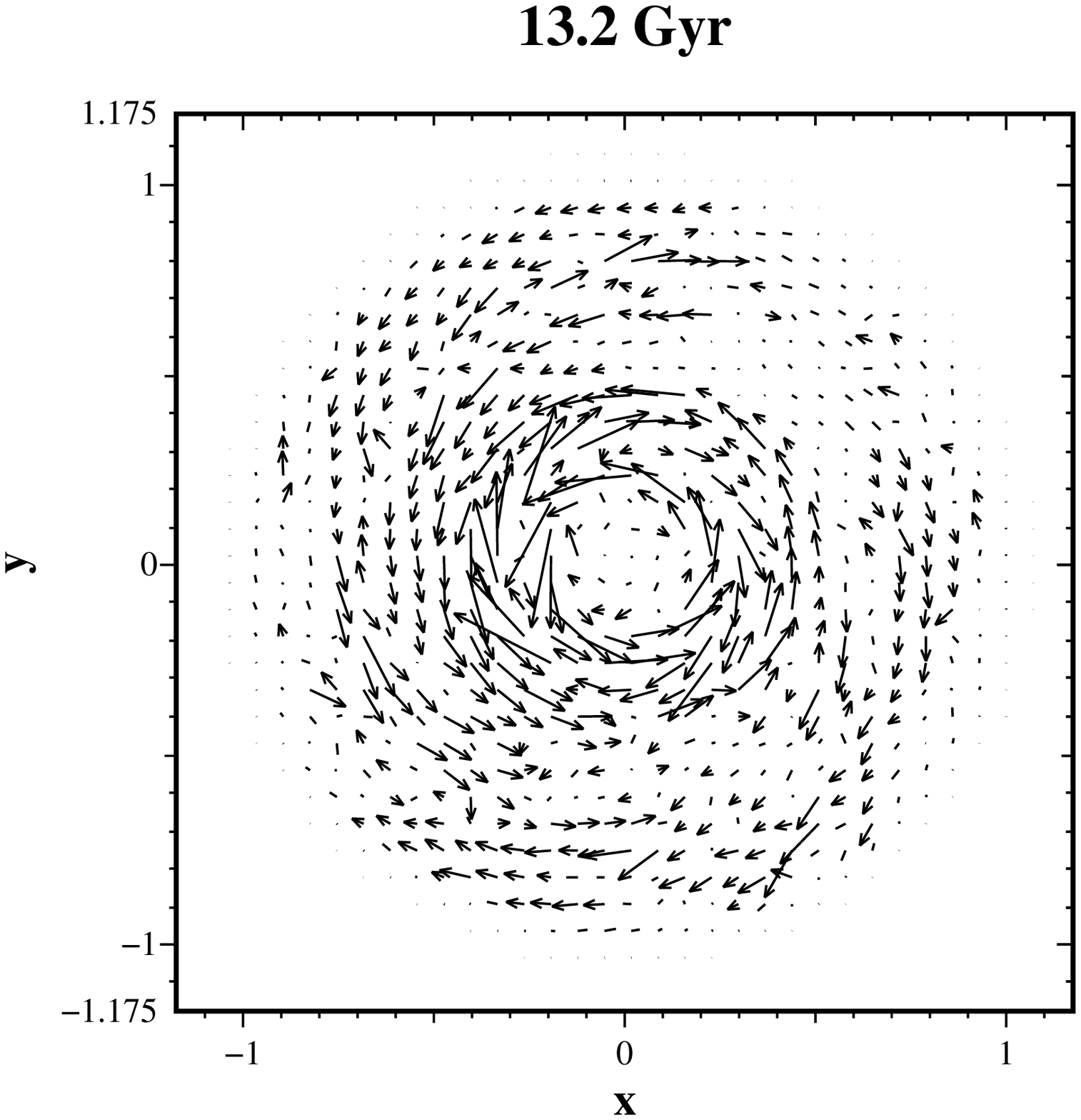}
\end{tabular}
\end{center}
\caption{Field vectors at galaxy ages $T=0.23, 0.78$\,Gyr (row (a)),
$T=2.3, 3.9$\,Gyr (row (b)), and $T=7.8, 13.2$\,Gyr (row (c)), for
the model with $R_\omega=20, R_\alpha=1, B_{\rm inj}=1$. The vectors
again give the magnetic field direction, their lengths are
proportional to the magnetic filed strengths. There is
one large-scale reversal visible at the present day ($T=13.2$\,Gyr).
(Model 135)}\label{fig:model135}
\end{figure*}

\begin{figure*}
\begin{center}
\begin{tabular}{ll}
(a)\includegraphics[width=0.41\textwidth]{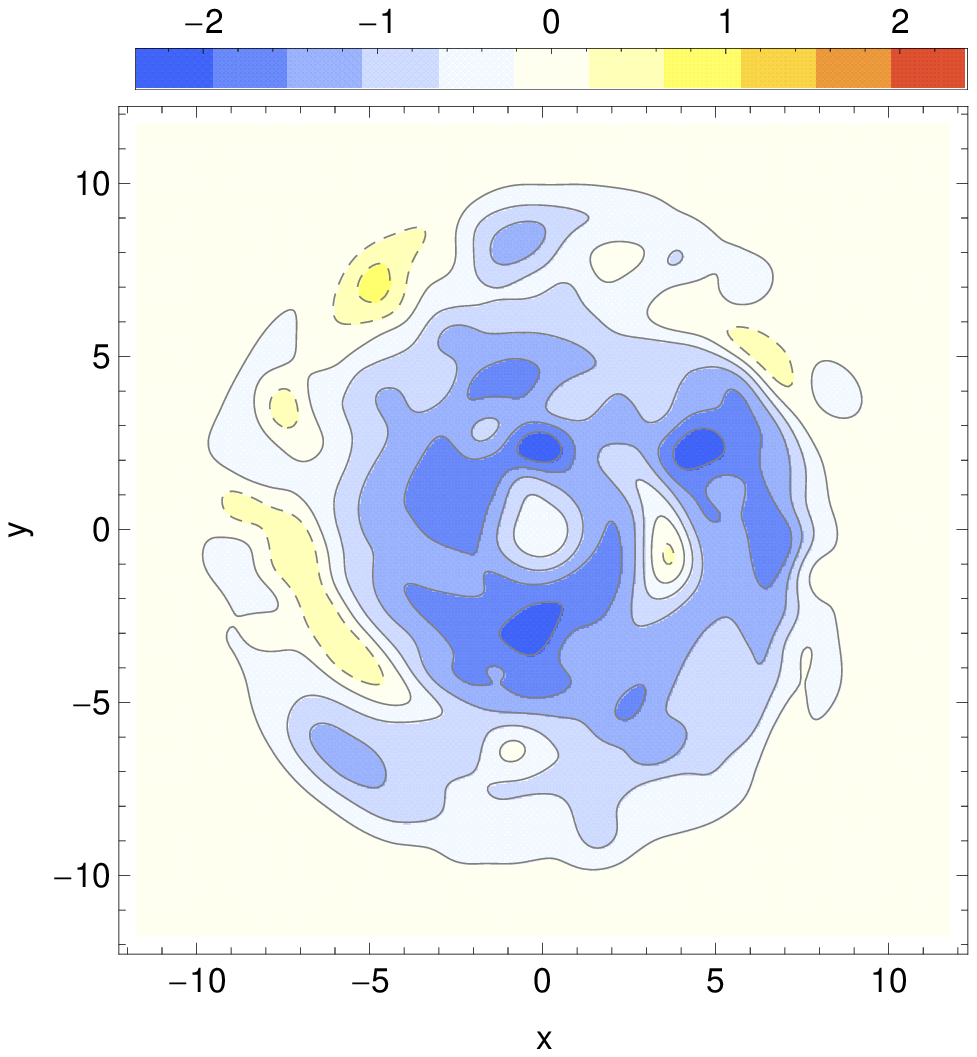}
 (b)\includegraphics[width=0.41\textwidth]{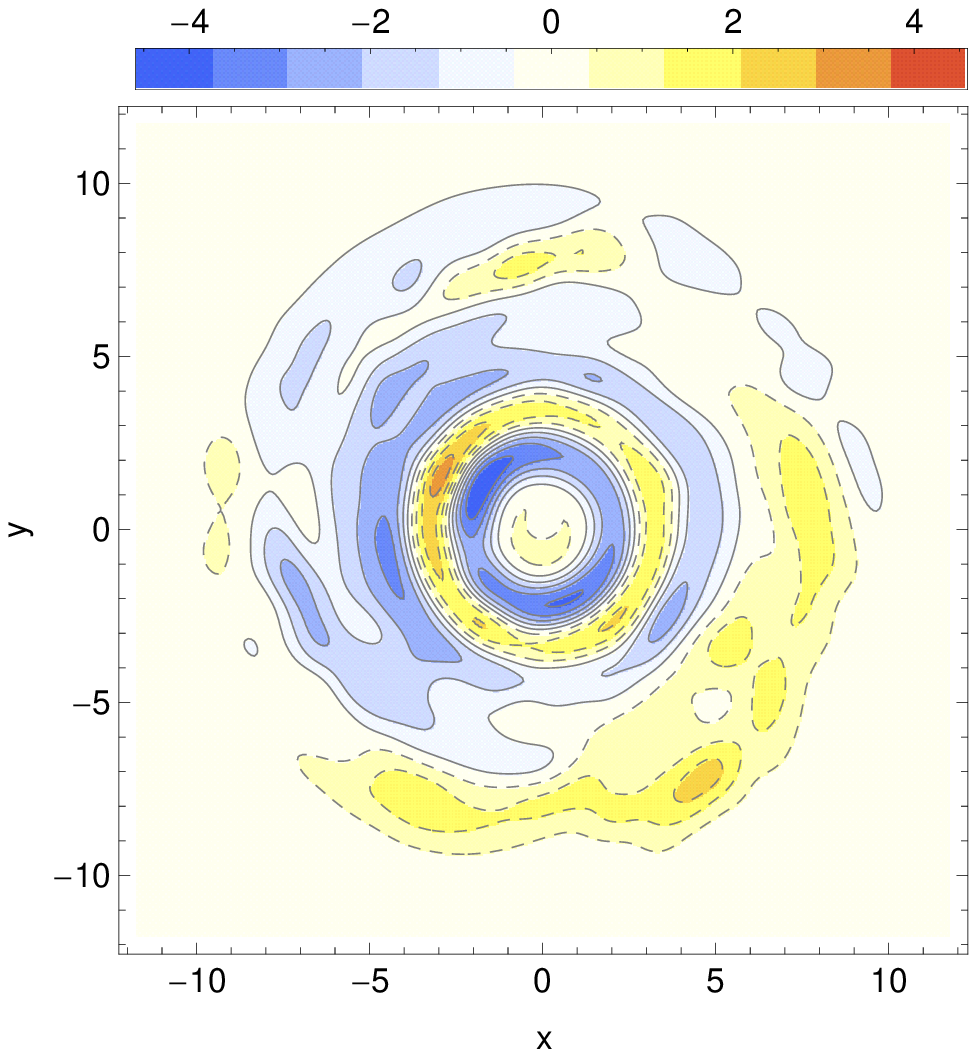}
\end{tabular}
\end{center}
\caption{Contours of azimuthal field (in units of $B_{\rm eq}$), smoothed by a
Gaussian filter of half-width 100\,pc, at galaxy age $T=13.2$\,Gyr.
Positive values (clockwise rotation) are indicated by continuous
contours, negative values (counterclockwise rotation) by broken
contours. (a) $B_{\rm inj}=1, R_\omega=10$, (b) $B_{\rm inj}=1,
R_\omega=20$.} \label{Bazimuth}
\end{figure*}

We discuss two models in some detail. They each have the standard
parameters governing field injection as discussed above, and have
$R_\alpha=1$, $B_{\rm inj}=1$. The first has $R_\omega=10$  (referred to as Model 138), and the
field at galaxy age $T=13.2$\,Gyr (present epoch) is seen to have no
large-scale reversals -- its evolution is shown in
Fig.~\ref{fig:model138}. The second example has $R_\omega=20$  (Model 135), and
large-scale field reversals are visible from $T=2$\,Gyr
to $T=13$\,Gyr -- see Fig.~\ref{fig:model135}. In
Fig.~\ref{Bazimuth} we show contours of the azimuthal field smoothed by
a Gaussian filter of 100\,pc half-width at $T=13.2$\,Gyr, which
illustrate clearly where large-scale reversals occur. In
Fig.~\ref{fig:model138}, panel (c) ($T=13.2$\,Gyr), local reversals
are visible near radius
$r=0.85$ ($x\approx -0.5, y\approx 0.7$, $x\approx 0.3, y\approx 0.8$),
which disappear with smoothing to
100\,pc. Similar features can be found comparing the last panel of
Fig.~\ref{fig:model135} with Fig.~\ref{Bazimuth}b.
 Other field configurations (with, say, a different number
and position of reversals) are possible,
depending on the dynamo parameters and the details of the field
injection process, and we return briefly to this topic later.

\begin{figure}
\includegraphics[width=0.45\textwidth]{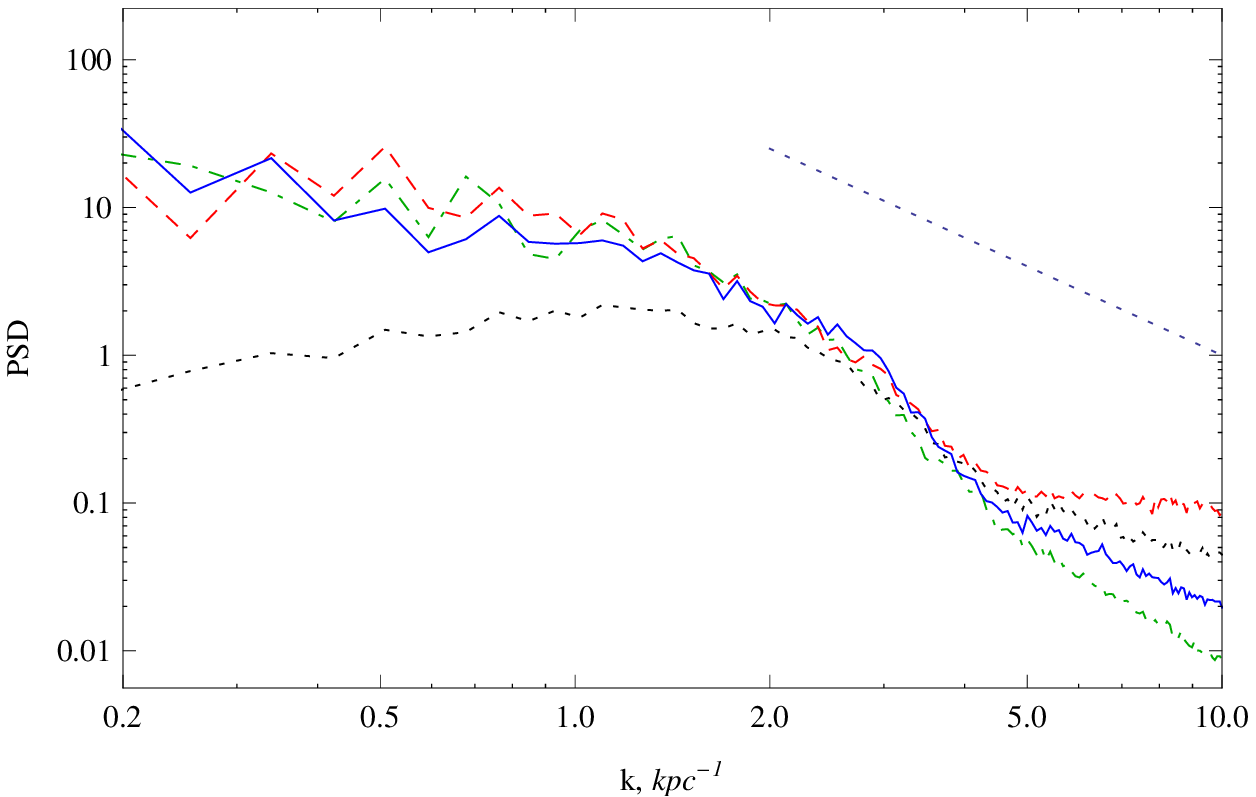}
\includegraphics[width=0.45\textwidth]{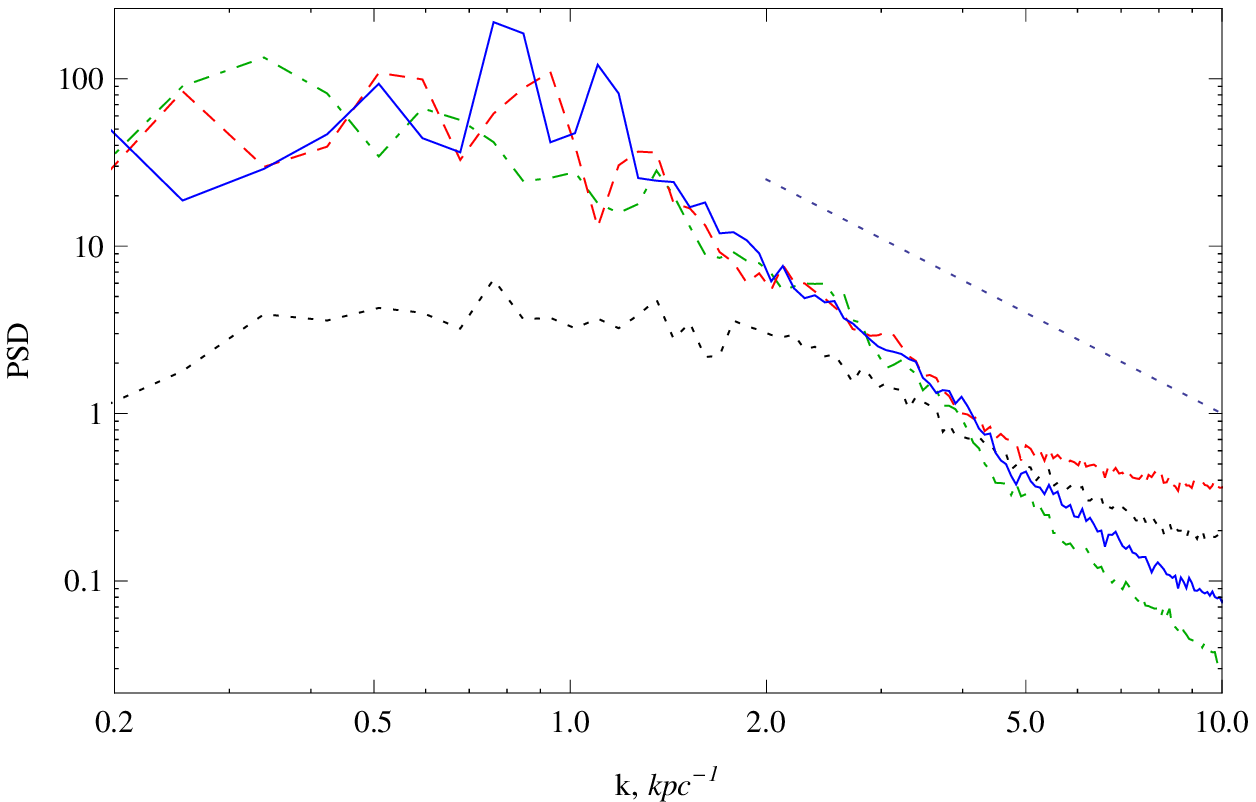}
\caption{Energy spectral density at time $t=0.3$ (dotted), $t=3$ (dot-dashed),
$t=10$ (dashed) and $t=17$ (solid). Upper panel:  $B_{\rm inj}=1, R_\omega=10$ (Model 138) , lower panel:
$B_{\rm inj}=1, R_\omega=20$ (Model 135). The broken line has slope $k^{-2}$.} \label{spec}
\end{figure}

\begin{figure}
\includegraphics[width=0.45\textwidth]{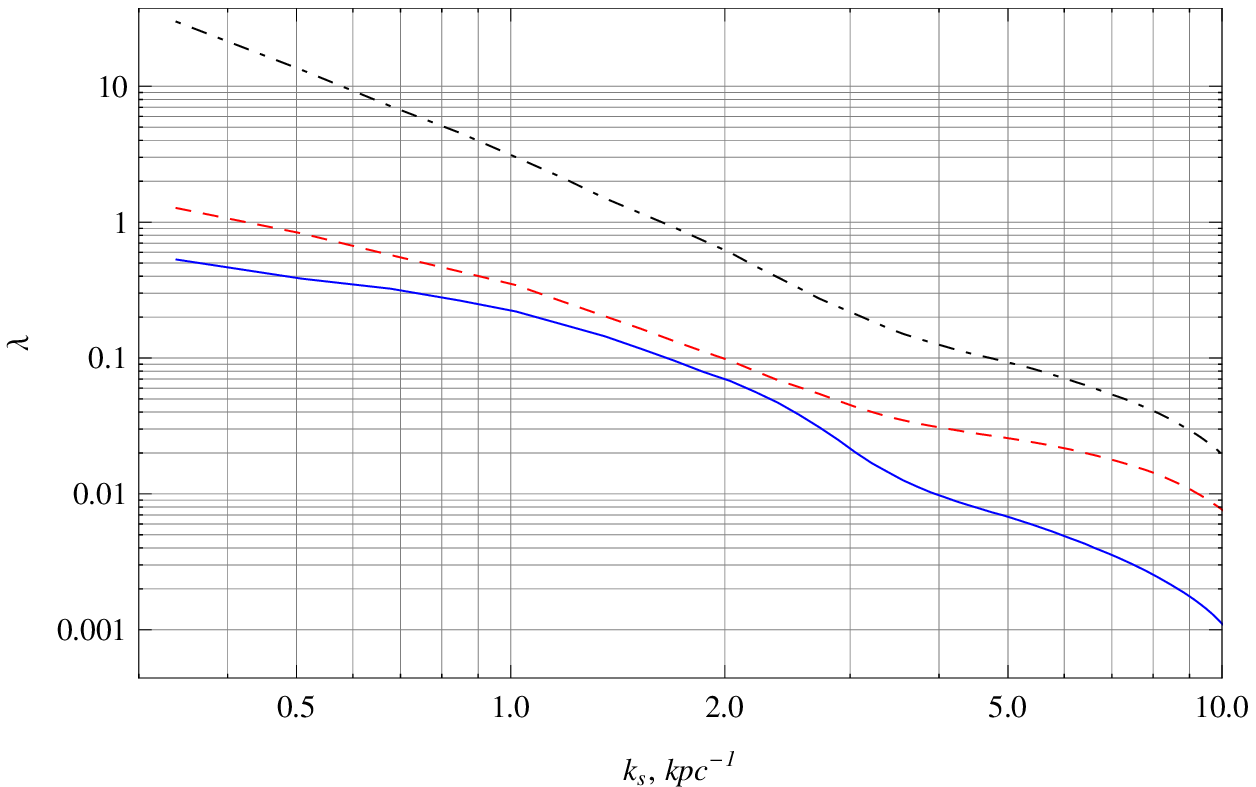}
\includegraphics[width=0.45\textwidth]{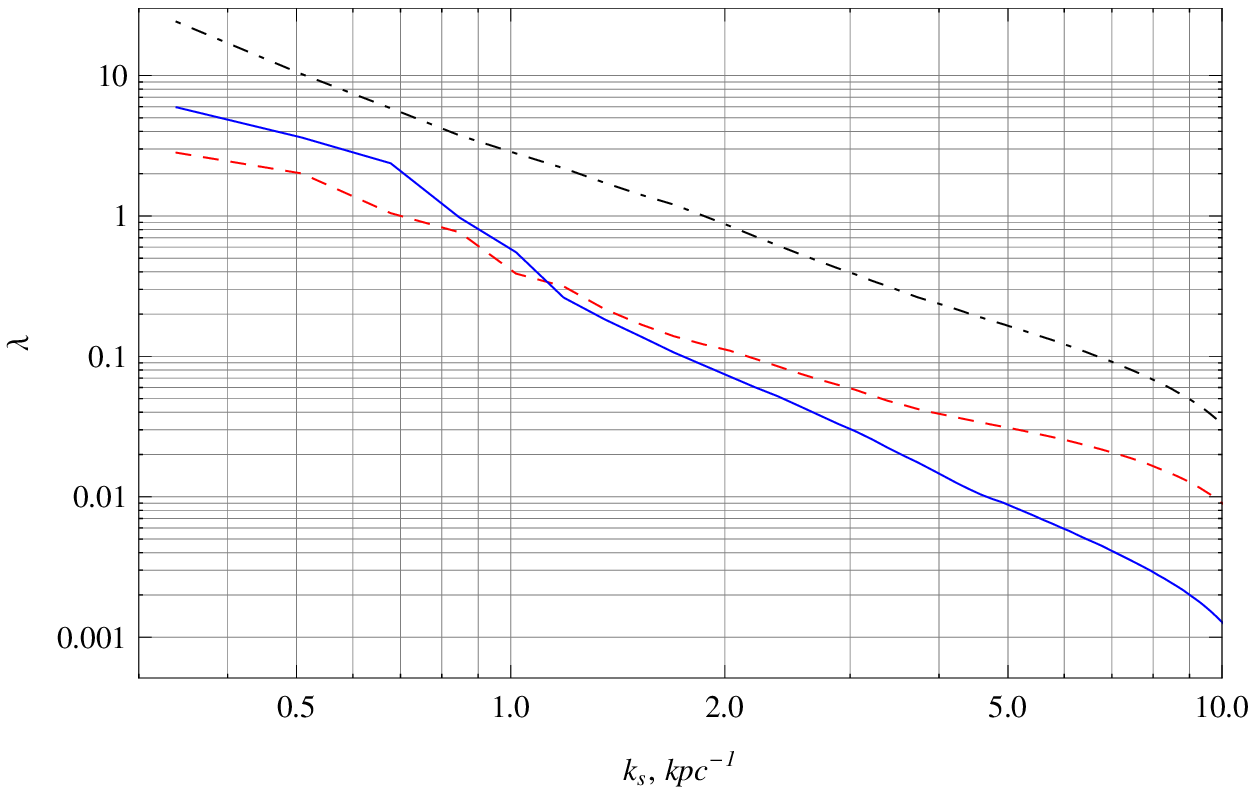}
\caption{Ratio of small to large-scale magnetic energies
$\lambda=\langle B(k>k_s) \rangle/ \langle B(k\le k_s) \rangle$ vs
scale separation parameter $\lambda$ at time $t=0.3$ (dot-dashed),
$t=10$ (dashed) and $t=17$: above - $B_{\rm inj}=1, R_\omega=10$ , below
$B_{\rm inj}=1, R_\omega=20$.} \label{rat}
\end{figure}

Typical energy spectra are shown in Fig.~\ref{spec} at several
times. The energy in the large-scale field ($k<5$) can be seen to increase with time, while that with $k>5$ decreases.
Model 135 shows an irregular distribution of energy in spectral space and large fluctuations in time.
The ratio of
small to large-scale magnetic field remains small, and decreases
with time --  see Fig.~\ref{rat}. In order to construct realistic
synthetic polarization maps it is necessary to complete the simulated
magnetic field by addition of an artificial turbulent contribution.
This is discussed in Sect.~\ref{sec:simul}.

An important result is that by the galaxy age $T=0.78$\,Gyr
(Figs.~\ref{fig:model138} and \ref{fig:model135}), a field of
strength $B_{\rm eq}$ with a scale of several kpc is already present. This
is a more-or-less inevitable result of the initially strong
small-scale fields being stretched by differential rotation and then
organized by the large-scale dynamo. We did also make a
cursory investigation of the effect of changes in
$R_\alpha$. Specifically, we looked at cases with $B_{\rm inj}=1$,
with $R_\omega=10$, $R_\alpha=2$ and $R_\omega=20$,
$R_\alpha=0.5$, i.e. the cases shown in
Figs.~\ref{fig:model138} and ~\ref{fig:model135} with
increased/decreased $R_\alpha$, respectively. The magnetic field
geometry is less sensitive to this parameter, and the
effect of these changes on field geometry is minor.
Increasing $R_\alpha$ in the first case increases the
regularity of the field at $T=$\sout{17}\, $13.2$\,Gyr ($t=17$),
and removes any
local reversals present. In the second case, the changes
were almost indiscernable. The global energy varies
slightly as the change in $R_\alpha$.

\subsection{The effects of varying the field injection parameters}
\label{changenspot}

\begin{figure}
\begin{center}
(a)\includegraphics[width=0.41\textwidth]{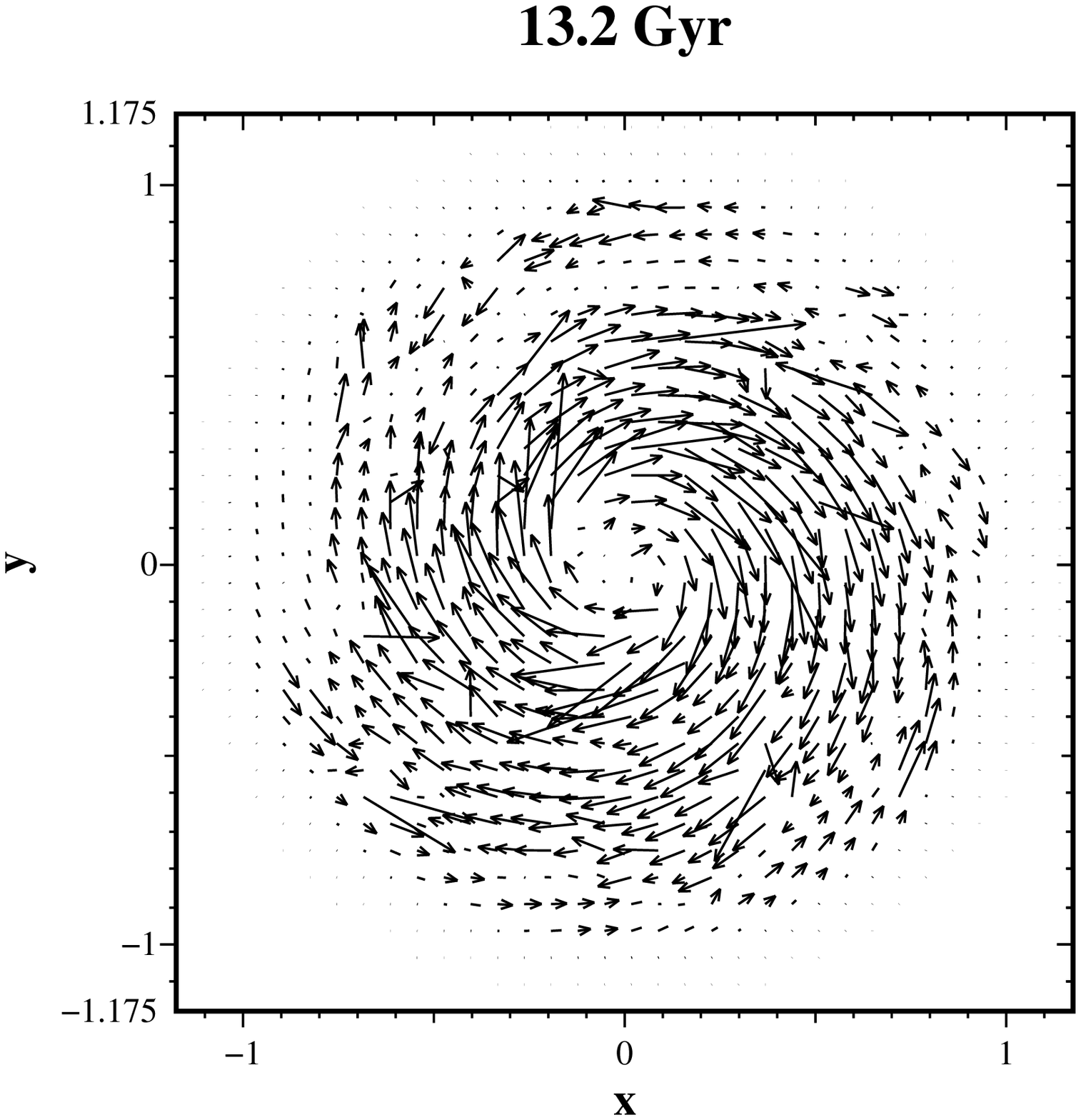}
(b)\includegraphics[width=0.41\textwidth]{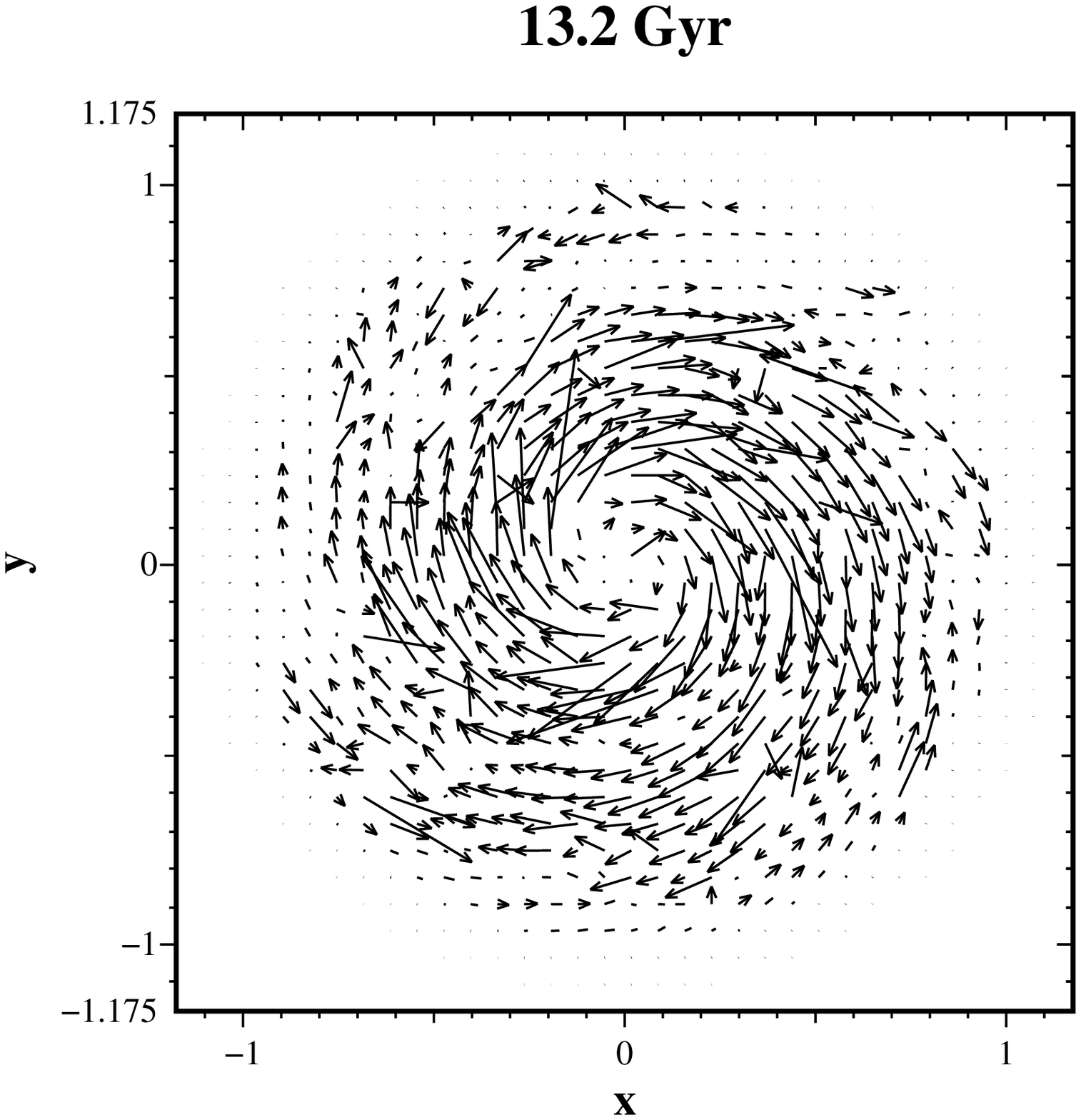} \\
(c)\includegraphics[width=0.41\textwidth]{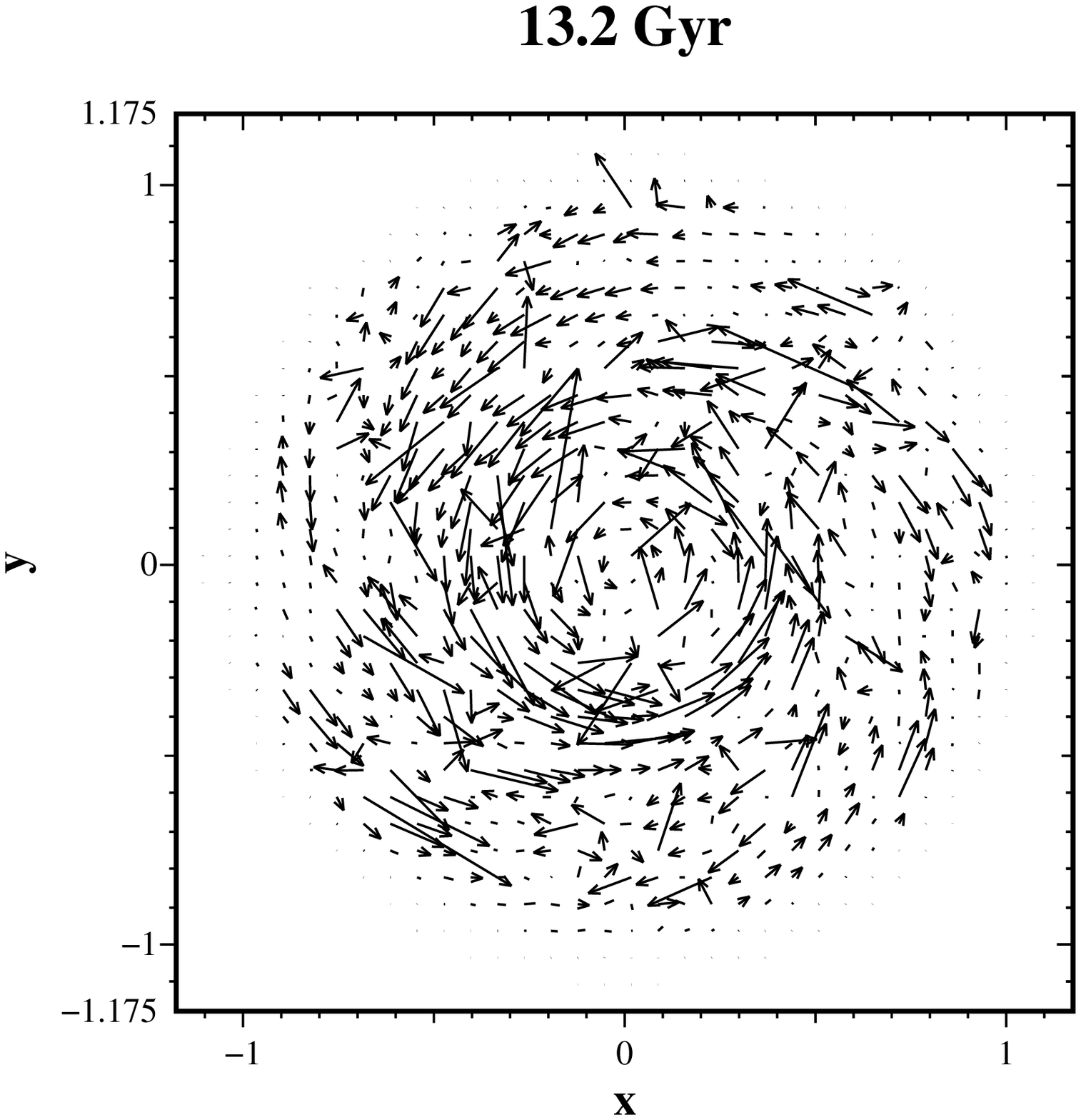}
\end{center}
\caption{To illustrate the effects of changing the parameters
$B_{\rm inj}$ and $n_{\rm sp}$. $B$-vectors at $T=13.2$\,Gyr for
models with $R_\omega=10, R_\alpha=1$, $n_{\rm sp}=50$ (cfed 100
normally); (a) $B_{\rm inj}=1$, (b) $B_{\rm inj}=2$, (c) $B_{\rm
inj}=4$.} \label{fig:nspoteq50}
\end{figure}

In order to demonstrate the roles of the parameters $n_{\rm sp}$ and
$B_{\rm inj}$, we show in Fig.~\ref{fig:nspoteq50} the field
configurations at $t=17$ (corresponding to an age of approximately
13.2\,Gyr) for models with $R_\alpha=1, R_\omega=10$ and $B_{\rm
inj}=1, 2, 4$ (i.e., increasing $B_{\rm inj}$ from the value used in
the models described in Sect.~\ref{changeromega}).

Reducing the value of $n_{\rm sp}$ to 50 reduces the amount of
disorder in the field (compare the last panel of
Fig.~\ref{fig:model138} with panel (a) of Fig.~\ref{fig:nspoteq50}).
As less small-scale field is being injected, this is perhaps not so
surprising.

We also now consider cases in which the turbulence in the
star-forming regions (spots) is assumed to be significantly stronger
than in the general ISM, giving an injected field larger than the
general equipartition field, i.e. $B_{\rm inj}>1$. Note that we do
not change the alpha-quenching formalism or the value of $B_{\rm
inj}$ in the spots when $B_{\rm inj}>1$, as the dominant dynamo
mechanism in the spots is assumed to be a small-scale dynamo to
which we cannot apply the alpha-quenching concept. Moreover the
filling factor of the high field region of the spots is small, and
the usual $\alpha\omega$ dynamo operates through nearly all the
disc. These assumptions are strongly supported by an experiment (not
shown) where with the parameters of panel (c) of
Fig.~\ref{fig:nspoteq50}, a {\it global} alpha-quenching
$\alpha=\alpha_0/(1+E)$, with $E$ a measure of the mean magnetic
energy over the disc, was used. The resulting field vectors were
very similar to those shown in panel (c) of
Fig.~\ref{fig:nspoteq50}.  As $B_{\rm inj}$ increases, obviously
does the disorder: compare panels (a), (b), (c) of
Fig.~\ref{fig:nspoteq50}. Further, the reduction of $n_{\rm sp}$
seems to produce some overall change to the field structure.
Specifically, the reduced value of $n_{\rm sp}$ means, quite
predictably, that for given $B_{\rm inj}$ the field is less
disordered. Less predictably, perhaps, even when $B_{\rm inj}$ is
increased and more disorder is apparent, the spiral structure is
better defined and rather more open, until $B_{\rm inj}=4$, see the
successive panels of Fig.~\ref{fig:nspoteq50}.

\begin{figure*}
\begin{center}
\includegraphics[width=0.49\textwidth]{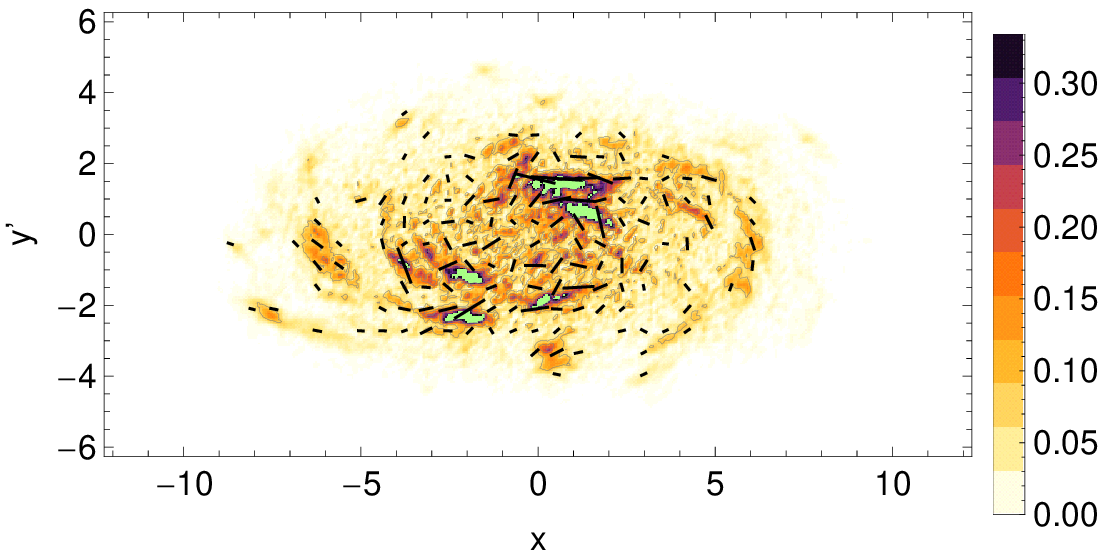}
\includegraphics[width=0.49\textwidth]{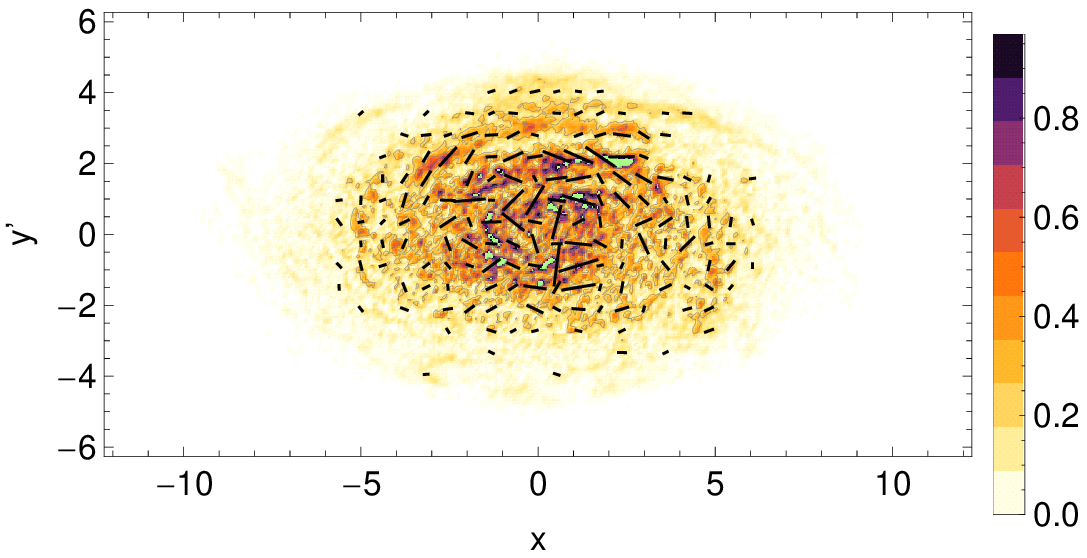}
\includegraphics[width=0.49\textwidth]{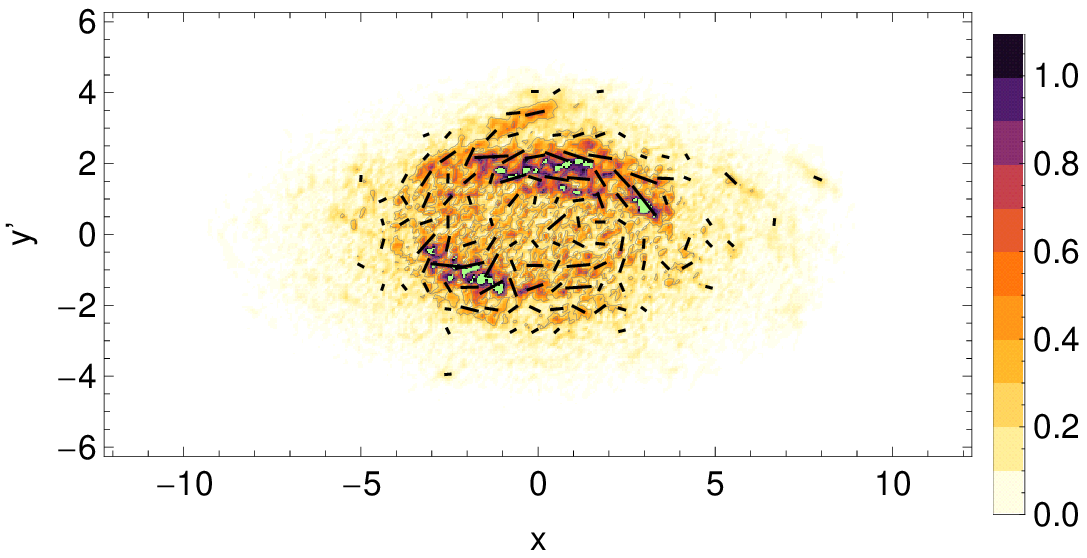}
\includegraphics[width=0.49\textwidth]{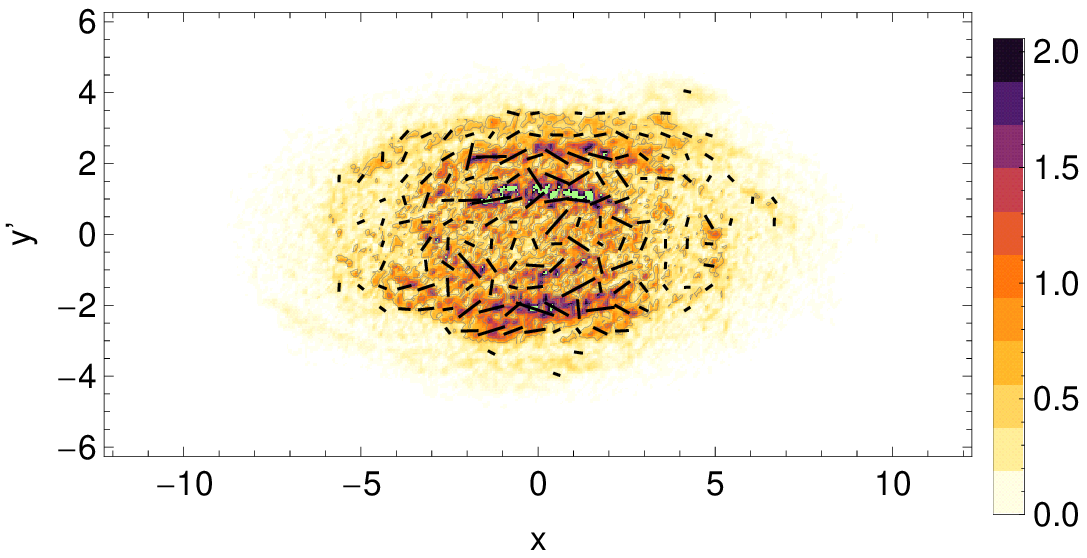}
\includegraphics[width=0.49\textwidth]{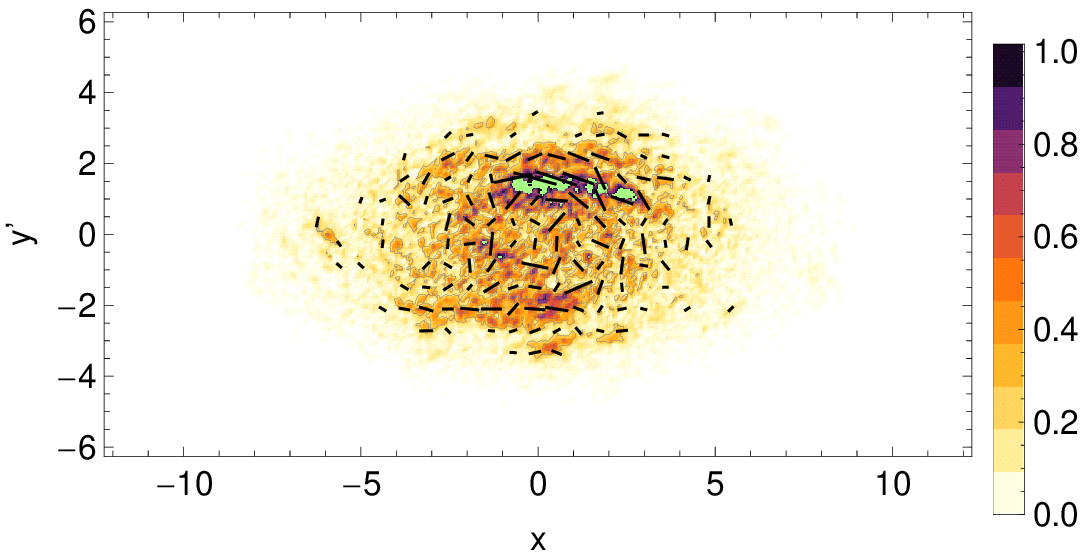}
\includegraphics[width=0.49\textwidth]{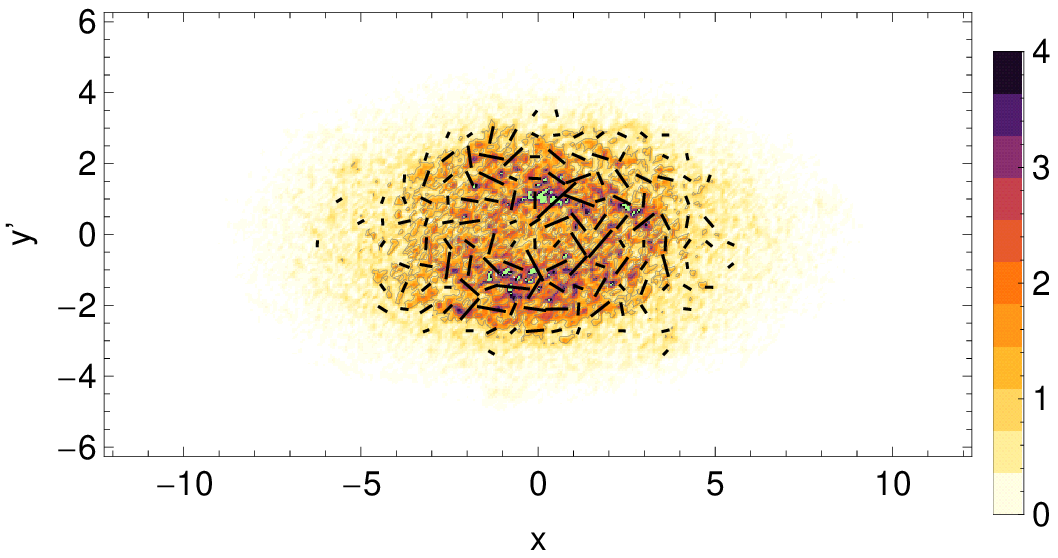}
\includegraphics[width=0.49\textwidth]{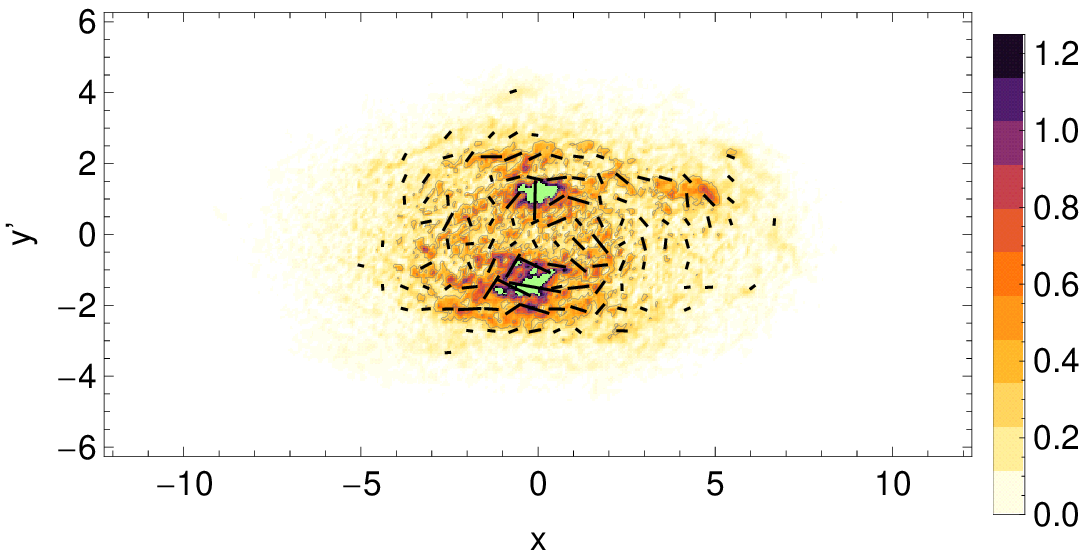}
\includegraphics[width=0.49\textwidth]{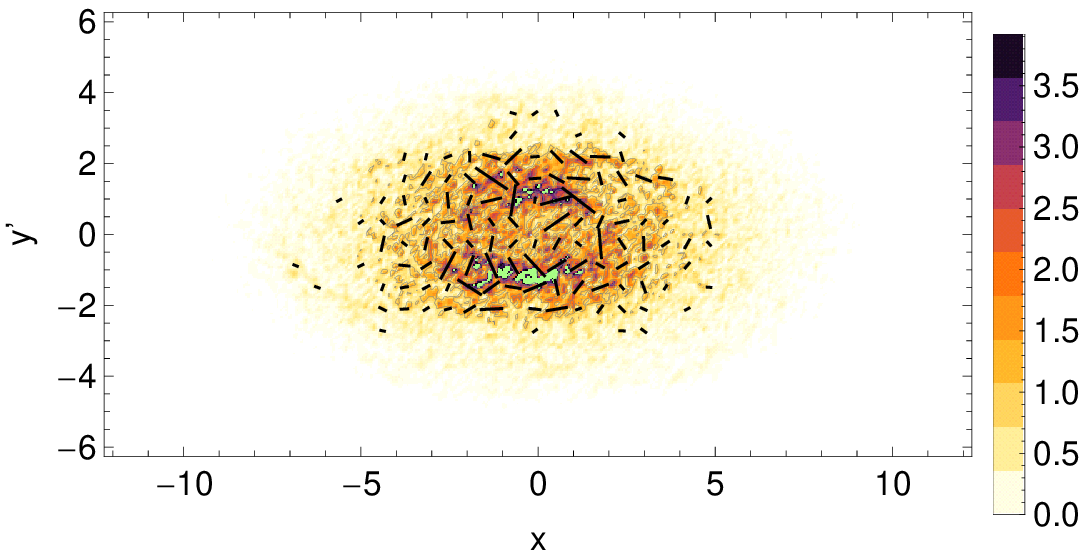}
\end{center}
\caption{Models 138 ($R_\omega=10$, left), 135 ($R_\omega=20$, right):
polarized synchrotron intensity (arbitrary units) at wavelength 21\,cm,
 as seen in the frame of the galaxy. Top to
bottom: $T=0.23$, $T=2.3$, $T=7.8$, $T=13.2$\,Gyr. The small bars show the (unsigned) field directions.}
\label{fig:pol_emis}
\end{figure*}

\begin{figure*}
\begin{center}
\includegraphics[width=0.49\textwidth]{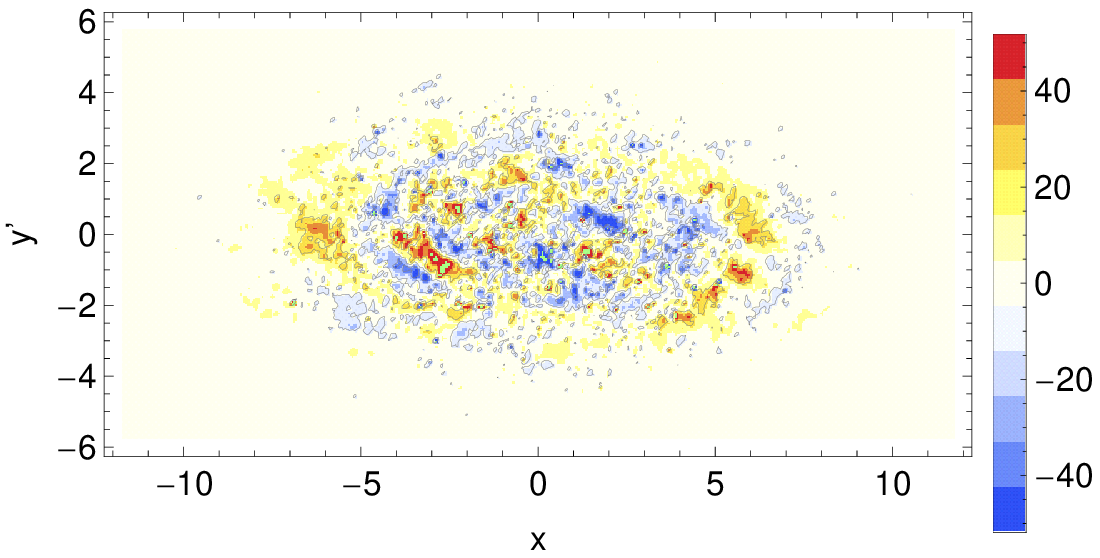}
\includegraphics[width=0.49\textwidth]{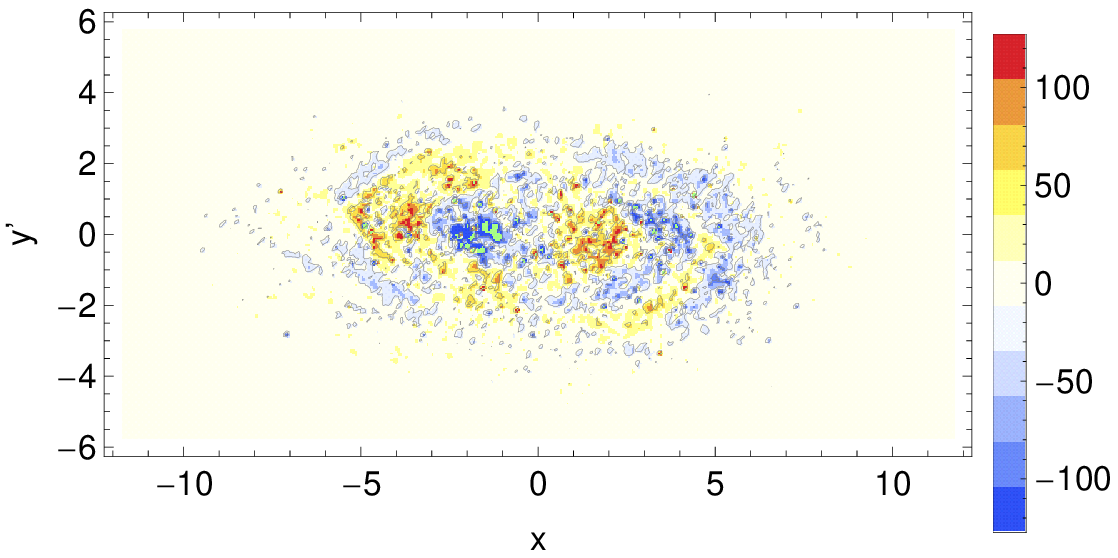}
\includegraphics[width=0.49\textwidth]{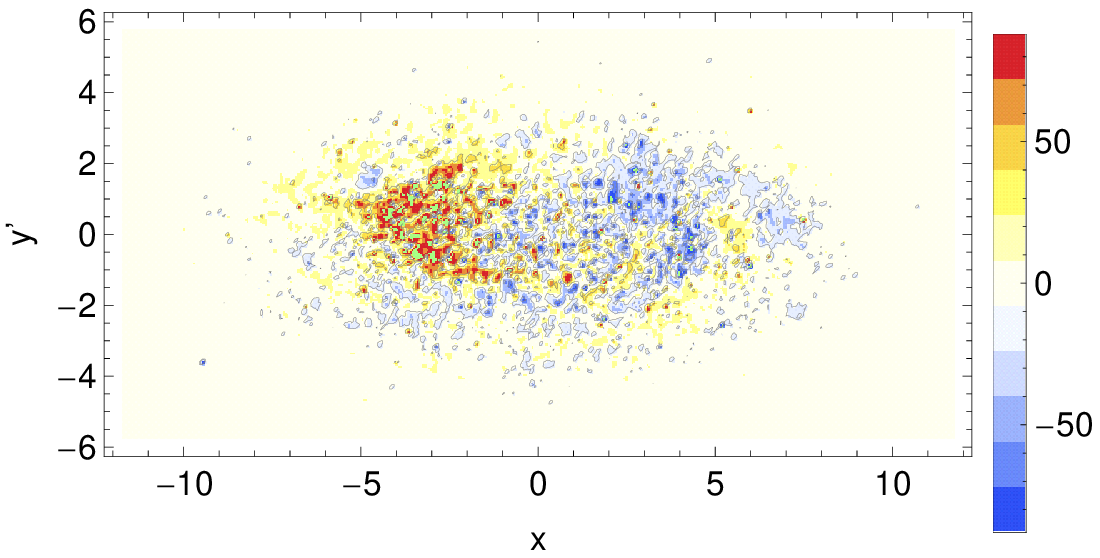}
\includegraphics[width=0.49\textwidth]{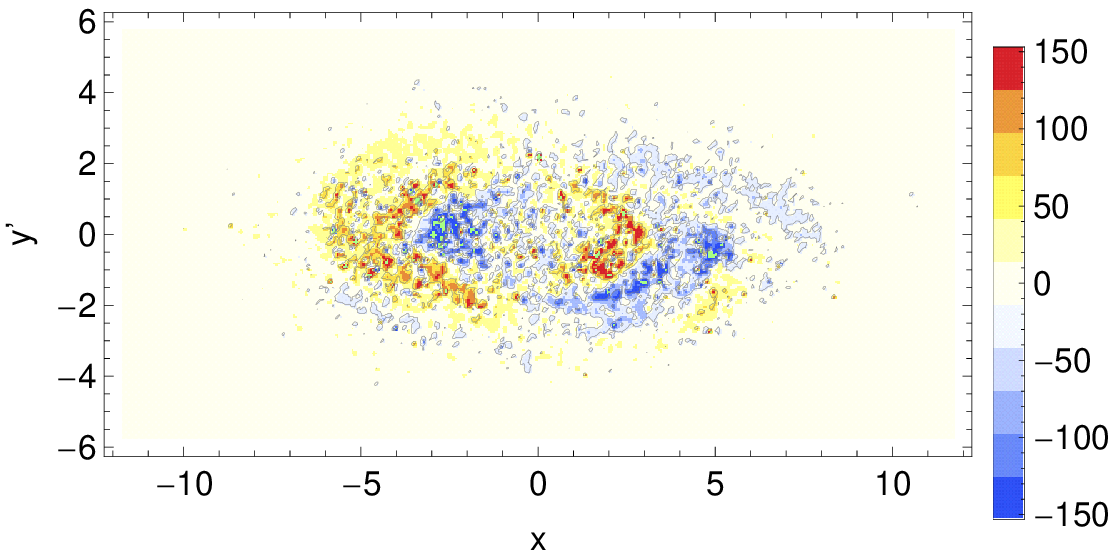}
\includegraphics[width=0.49\textwidth]{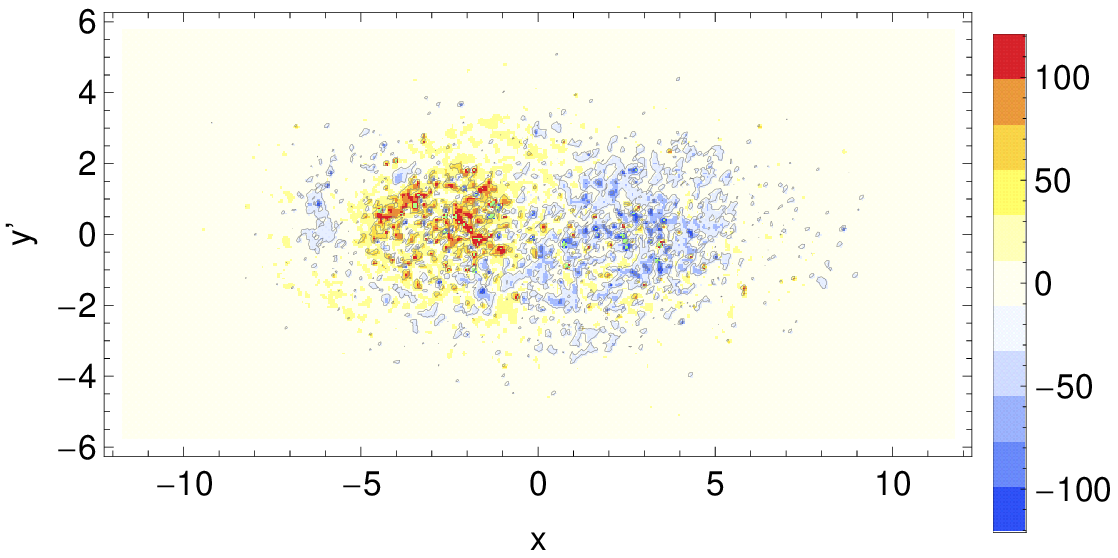}
\includegraphics[width=0.49\textwidth]{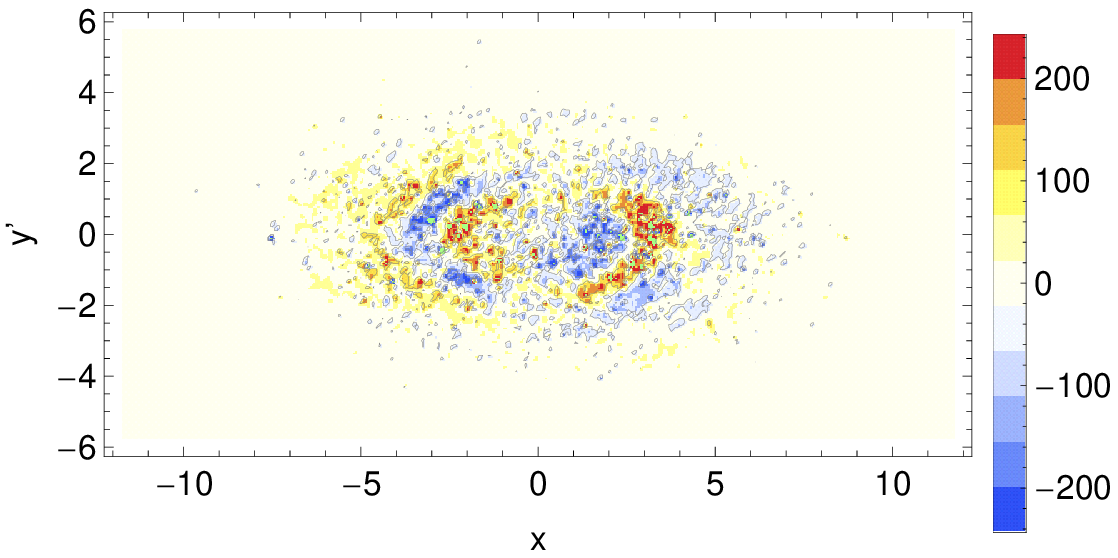}
\includegraphics[width=0.49\textwidth]{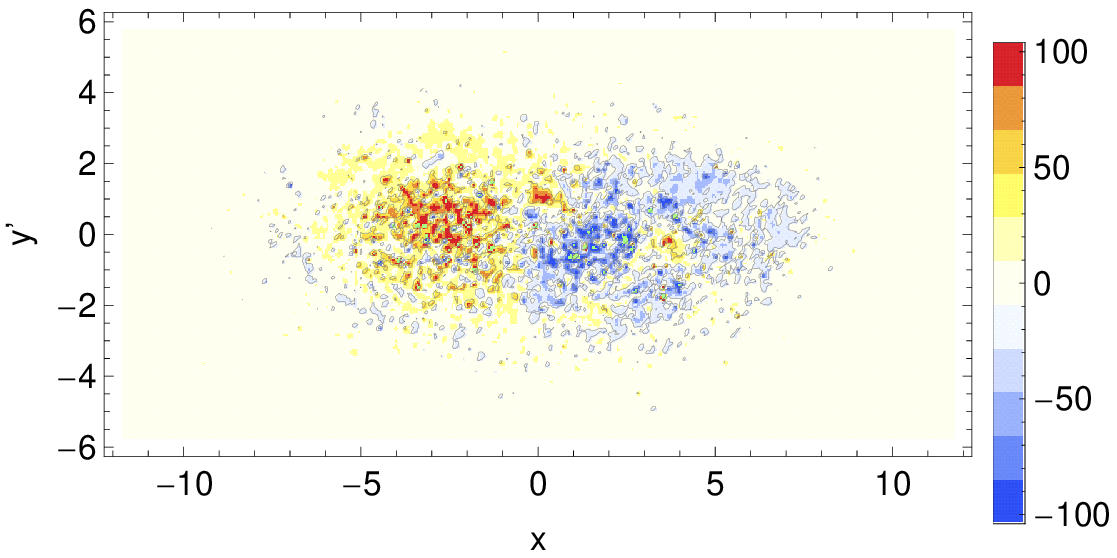}
\includegraphics[width=0.49\textwidth]{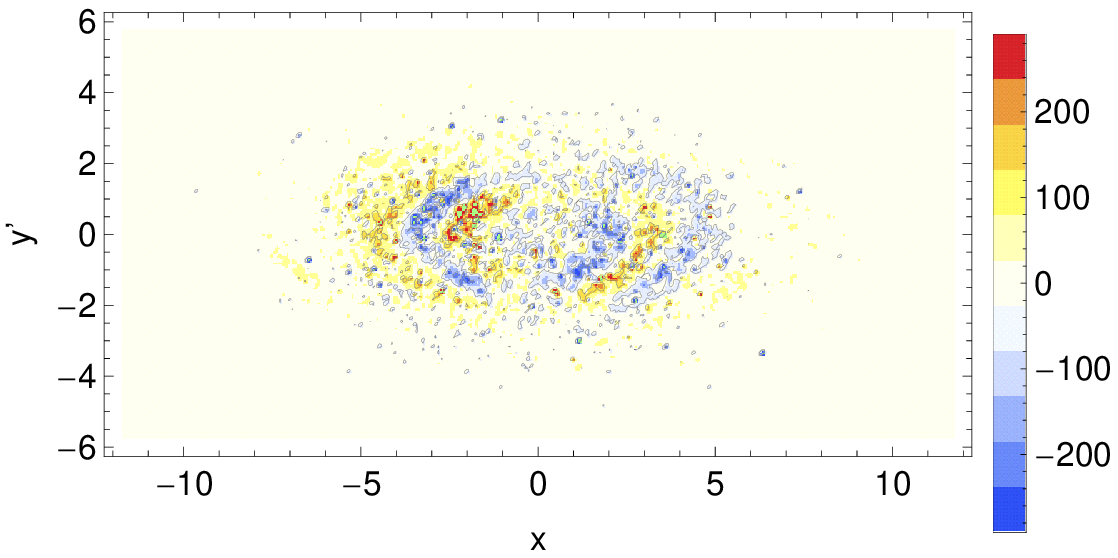}
\end{center}
\caption{Models 138 (left), 135 (right):
Faraday rotation measures (in $\rm rad/m^2$ assuming  $B_{\rm eq}$=1\,$\mu G$) at wavelength 21\,cm, as seen in the frame of the galaxy. Top to
bottom: $T=0.23$, $T=2.3$, $T=7.8$, $T=13.2$\,Gyr.} \label{fig:RM}
\end{figure*}

\section{Simulations of regular magnetic fields, total intensity,
polarization and Faraday rotation of evolving galaxies}
\label{sec:simul}

In order to construct an extension of the 2D magnetic
field ${\mathbf B}(x,y)$ of our models to a 3D distribution
${\mathbf B}(x,y,z)$ in a galactic disc, use  of a
Fourier transform technique seems natural.
We adopt the approach of writing the horizontal
magnetic field in factorized form as
\begin{eqnarray}
\hat{B}_x(k_x,k_y,k_z)=\hat{B}_x(k_x,k_y) \hat{g}(k_z a_h),\\
\hat{B}_y(k_x,k_y,k_z)=\hat{B}_y(k_x,k_y) \hat{g}(k_z a_h),\\
g(z)=\exp(-z^2), \label{fbxy}
\end{eqnarray}
where $(\hat.)$ denotes the Fourier transform, ${\mathbf k}$ is the
wave vector and $a_h$ is a vertical scale. The structures in the
horizontal plane are extrapolated in height with the same scales
as the associated horizontal structure,
$a_h=(k_x^2+k_y^2)^{-1/2}$. We limit $a_h$ by the galactic scale
height $h_g=2$\,kpc. The solenoidal condition allows the computation of the
vertical magnetic field component
\begin{eqnarray}
\hat{B}_z(k_x,k_y,k_z)=-\frac{\hat{B}_x(k_x,k_y,k_z)k_x+\hat{B}_y(k_x,k_y,k_z)k_y}{k_z}.
\label{fbz}
\end{eqnarray}
The turbulent component of magnetic field in the whole galactic disc is
evaluated in the same manner as suggested by
\citet{2011AN....332..524A}. The thermal electron distribution is assumed as
\begin{equation}
n_e(r,z)=n_0\exp{\left[-\left(\frac{r}{r_{0}}\right)^2\right]}\exp{\left[-\left(\frac{z}{h_0}\right)^2\right]},
\end{equation}
where exponential scales are $r_0=5h_0=5.9$\,kpc and
$n_0=0.03$\,$\rm cm^{-3}$. The cosmic-ray density has the same shape
but measured in an arbitrary units. The synchrotron spectral index
is 0.8.

The time evolution of the total and polarized synchrotron intensity
and RM in the rest frame of the galaxy inclined at 60 degrees for the
models 138 and 135 is shown in Figs.~\ref{fig:pol_emis} and
\ref{fig:RM} as a series of snapshots. Here we take $B_{\rm eq}=1\mu{\rm G}$.

Polarized emission is detected already in the early phases of galaxy
evolution around galaxy age $T=0.23$\,Gyr, corresponding to redshift
$\approx7.5$ (Fig.~\ref{fig:pol_emis}, first row), but the intensity
is still low. The equipartition level is reached around $T=1$\,Gyr
(redshift $\approx4$), so that strong emission is seen after
$T=2.3$\,Gyr (Fig.~\ref{fig:pol_emis}, rows 2--4). The two models
show different distributions of the polarized emission across the
galaxy, but these will depend slightly on the choice of the initial
parameters. The occurrence of large-scale field reversals, the main
difference between the models with different choice of the dynamo
number $R_{\omega}$, cannot be distinguished from
Fig.~\ref{fig:pol_emis} because polarized emission is not sensitive
to field reversals.

Faraday rotation measures (RM) (Fig.~\ref{fig:RM}) clearly show the
development of field reversals with galaxy age. At early times, RM
is significant in restricted areas. Large-scale RM patterns are seen
after age $T\approx2$\,Gyr (redshift $\le 3$), in excellent agreement with
the estimate given by Arshakian et al. (2009). The main result of
the improved model of this paper is the occurrence of large-scale
reversals of Model 135 (Fig.~\ref{fig:RM}, right column), which
remain clearly visible until the present epoch.

Note that the polarized intensity and RM values in
Figs.~\ref{fig:pol_emis} and \ref{fig:RM} refer to the galaxy frame.
These maps will look different in the rest frame of the observer
because of a frequency shift. Highly redshifted galaxies emit at
high frequencies at which the depolarization effects are different:
depolarization becomes weaker with increasing frequency. Weak
regular magnetic fields at high redshifts and strong dilution of RM
by a factor of $(1+{\sf z})^{-2}$ (here $\sf z$ is redshift) lead to
small observable RM for distant galaxies. The effect of
depolarization for polarized intensity and RM maps is discussed in
Arshakian et al. (2011). Detection of RM patterns from distant
galaxies will become possible only for large forthcoming radio
telescopes such as the SKA.

\section{Discussion}
\label{sec:discussion}

We can identify two stages in the development of the magnetic field
from the random seed field to a mature magnetic field configuration.
In the first stage the magnetic field has a spatially intermittent
structure, whilst in the second the mature magnetic field has a
regular structure with a large-scale component which is an almost
axisymmetric spiral. The transition time from the first to the
second stage is quite robust and depends on the diffusion time $t_d$
of the magnetic field through the vertical extent of the galactic
disc ($t/t_d \approx 2$) which corresponds to 1--2\,Gyr in
dimensional units. If we assume that the galactic disc was formed at
redshift around $10$, the transition to the mature field occurs at
redshift $\approx 4-3$, which is too early to be observed even by
the SKA. If however a galactic encounter or strong interaction
'resets the clock', and so can be considered as the starting point
for the field evolution, observation of a spotty field configuration
or of a large number of reversals appears a realistic possibility.

The SKA will be able to recognize the regular field structures of
nearby galaxies from RMs of background polarized sources behind a
galaxy and continuous RM map of the diffuse polarized emission of a
galaxy (Stepanov et al. 2008). The former approach will allow the
simple field structures to be recognized at a distance of about 100
Mpc in disc galaxies having a  size of $20$\,kpc. With SKA angular
resolution of 1\,arcsec at 1.4\,GHz, mapping of the diffuse emission
of a galaxy itself can resolve RM patterns up to a distance $z
\approx 0.11$, which corresponds to a comoving radial distance of
$\approx 450$\,Mpc.

We also learn from this work that the magnetic field evolution
continues slowly over a very long time after a mature field
configuration first appears. This evolution affects the size and
exact location of the radial range (``ring'') filled with a spiral
field of a given direction with respect to azimuth.

We also considered purely regular initial seed fields and also weak
random initial seed fields with no injection, and confirmed that
both evolve into smooth large-scale fields.

What is quite new and unexpected is the possibility to obtain
several (two or even three) rings of oppositely directed magnetic
fields in neighbouring rings and hence global field reversals
between $T\approx3$\,Gyr and 13\,Gyr. Global field reversals
are not observed yet in nearby external galaxies. A reversal of
several kpc extent in azimuth has been found between the Orion and
Sagittarius arms in the Milky Way from pulsar RM data (e.g. Frick et
al. 2001, Noutsos et al. 2008, Nota \& Katgert 2010, Van Eck et al.
2011). Another possible reversal between the Orion and Perseus arms
is difficult to prove or disprove with the data currently available
(see Frick et al. 2001). Han et al. (2006) claimed that many
magnetic field reversals can be recognized in the Milky Way. 
However, the existence of multiple reversals is not statistically
well established (Men et al. 2008). The global field structure of the Milky Way
it is not known yet.

Standard galactic dynamo models which start from a very week seed
field give usually a final magnetic field configuration without
reversals. Traditional mean-field dynamo theory has considered
reversals as long-lived transients (Vasil'yeva et al. 1994, Moss et
al. 1998, 2000, Petrov et al., 2001, 2002) which occur if the
initial field is strong enough and has a suitable configuration.
These traditional 1D models are however obviously too oversimplified
to be fully convincing because they assume {\it inter alia}
that the galactic magnetic field is purely axisymmetric. Here we
consider a 2D model where the magnetic field can have an arbitrary
configuration. We consider the results obtained in this paper to be
sufficiently detailed to produce reliable results on the occurrence
of reversals.

The strongly nonlinear initial conditions of our models (with field
strengths of order the equipartition strength, even though with a
small volume filling factor and no large-scale order) play a key
role in determining the long-term evolution and structure of the
magnetic field. Unlike conventional dynamo models, which begin with
a weak seed field often of large scale, with no
further intervention, the saturated state is now not solely
determined by the dynamo parameters and the galaxy structure: the
presence and nature of the small-scale magnetic
field injections, especially that at $T=0$  is crucial. However, the
dynamo number $R_\alpha R_\omega$ is also important: We show in
Figs.~\ref{fig:model138} and \ref{fig:model135} the differing field
geometries obtained when $R_\omega$ is doubled. In the first case
there are no large-scale field reversals at $T=13.2$\,Gyr, but in
the second a large-scale reversal is present. We found other models
with two reversals present. Briefly, a larger dynamo number and a
larger value of $B_{\rm inj}$ favor the presence of field reversals,
cf. Figs.~\ref{fig:model138}, \ref{fig:model135} and
\ref{fig:nspoteq50}. We intend to discuss this interesting aspect of
the solutions in more detail in a separate paper (Moss \& Sokoloff,
in prep.).

We stress that a relatively small variation of the dynamo-governing
parameters converts a model without global field reversals into a
model with one or several global reversals.  One important
parameter turns out to be the dimensionless
number $R_\omega$ which is
determined by the angular rotation velocity, the galaxy size and the
turbulent diffusivity. Faster galactic rotation tends to a larger
value of $R_\omega$ and hence to a higher probability to generate
one or more long-lived field reversals, such as that detected in the
Milky Way. However, the Milky Way is not rotating faster than nearby
spiral galaxies, where no large-scale reversals have been found (e.g. M31,
M83, NGC\,6946; see Beck 2010).

Even if the rotation curve, the disc thickness and the $\alpha$
coefficient could be measured with high accuracy, the value of the
turbulent diffusivity will remain uncertain. Direct numerical
simulations of particular models of galactic hydrodynamics should
investigate the question of global reversals, to support (or reject)
the results from this paper.

We can hope to calibrate our model by comparison of our
'final' state at $T\approx 13.2$\,Gyr with the fields of
well-observed nearby galaxies. If there is reasonable
agreement, this might suggest that the historical
evolution of the models is reasonably valid.

The particular small-scale seed
field configuration (and injection process) in a given galaxy is
obviously unknown. We predict that some galaxies are expected to have mature
magnetic fields without reversals whilst others will have  a small
number of field reversals.
Thus it becomes increasingly important to amass a large number of
high-quality radio maps of these objects that are capable of
unambiguously detecting field reversals. Then the statistics of
observed reversals can be compared with those of our model.

Given that the number of global reversals seems to be a
suitable indicator for the magnetic field configuration
and that this is relatively easy to determine
observationally and to interpret theoretically, our
suggestion is to resolve the problem by observations,
i.e. to determine the number of global reversals in
various morphological types of young and mature galaxies.
This quantity could then be used to constrain the
quantities  describing the galactic hydrodynamics.

The model presented in this paper could be exploited in two ways, to
simulate the magnetic field evolution from disc formation, as well
as to understand the variety of magnetic field configurations in
contemporary galaxies. The latter problem arises from the major
controversy that observations revealed at least one global field
reversal in the Milky Way, but not in other well-investigated
nearby galaxies, in particular in the Andromeda galaxy M31, although
the available spatial resolution is sufficiently high in this case.
Several solutions are viable. It might be claimed that
large-scale reversals are rare and it is by chance that
they are present in our own Galaxy. Other examples of this kind may be
unknown because of poor statistics. The available set of
high-quality data for nearby galaxies is possibly too small to
isolate field reversals. Another way of thinking is to claim that
the dynamo in M31 only works in the ring of star formation at about
10\,kpc radius and so cannot produce several rings with opposite
field directions. Future observations with the SKA and its precursor
telescopes are expected to enlarge substantially the data set and to
resolve this controversy.

In order to be simple and specific, our model ignores several
important features of the galactic dynamo which might determine the
final dynamo-generated configuration of the large-scale magnetic
field. In particular, we have ignored problems connected with the
impact of galactic winds on the galactic dynamo (see e.g. Moss et
al. 2010, Dubois \& Teyssier 2010) or of the accretion of external
fields into the galactic dynamo (Moss \& Shukurov 2001). Another
important issue is the role of magnetic helicity conservation in
nonlinear suppression of the galactic dynamo (Shukurov et al. 2006,
Sur et al. 2007). Further simulations would be desirable here. Our
model is valid for thin disc galaxies and cannot be used for
dwarf and irregular galaxies. Another steps to improve the model is
to take into account the suppression of the synchrotron emission by
inverse Compton losses off the cosmic microwave background at high
redshifts, cosmological "downsizing" of galaxies, and X-shaped halo
fields (for details see Arshakian et al. 2011).

We have made several attempts to generalize our model and include
various secondary effects. For example, we investigated the
possibility that the $\alpha$ coefficient as a dynamo-governing
quantity can have (at least at the early stages of galactic
evolution) a ``spotty'' structure. If we adopt a ``spotty $\alpha$''
model, then the effective value of $\alpha$ averaged over the disc
is reduced by the volume filling factor of the spots. Thus to get
$R_\alpha R_\omega=O(10)$ needs unrealistically large values of
$R_\alpha$ (or $R_\omega$), so that this model seems implausible.
Also, such a model does not produce significant smaller-scale
structure. Correspondingly, we do not consider models with ``spotty
$\alpha$'' as a reasonable way to explain the observed magnetic
fields of galaxies.

A further option presented in the literature is to start with a
strongly organized but dynamically weak intergalactic field at
very early times originating, say, from a purely homogeneous
cosmological magnetic field and to ignore further intervention of
small-scale magnetic fields. The traditional opinion here (e.g. Beck
et al. 1996) is that the galactic dynamo picks up such field as a
seed and over a timescale of some Gyrs approaches a final
magnetic field configuration. We explored this option in the framework of
our model which yielded an unexpected result. A large-scale seed
field having a symmetry (dipole symmetry in form of a dipolar
perpendicular to the rotation axis) that is different from the
expected final configuration (axisymmetric quadrupolar
symmetry) first decays to negligible strengths and then grows in to
the final quadrupolar configuration. (In contrast to F.\,Krause \&
Beck (1998) who stressed that the efficient growth of quadrupolar fields
needs quadrupolar seeds.) This evolution takes a long time, that
comfortably exceeds the galactic age.
We will present a description of this unexpected result in a separate paper
(Sokoloff \& Moss, in preparation). If we take the small-scale
magnetic injection into account, the magnetic field grows and
reaches the final stage in a reasonable time.  Indeed,
equipartition strength fields of kiloparsec scale are present after
1--2\,Gyr. The point is that the field configuration at large times
is determined by the injections as well as the seed field.

\section{Conclusions}
\label{concl}

We have adopted a new approach to the evolution of
dynamo-generated magnetic fields in spiral galaxies.
Our attempt to model the interaction of fields generated by
small-scale dynamo action in discrete star forming regions together
with a global-scale dynamo takes us beyond standard mean field
theory. Our mechanism for the introduction of small-scale fields is
necessarily rather arbitrary, but we feel it illustrates an
important physical mechanism.

Importantly, the models robustly possess large-scale fields
of about equipartition strength on scales of several
kiloparsecs after 1--2\,Gyr (i.e. at redshift $\sim 4-3$).

Our numerical experiments demonstrate that the large-scale dynamo is
indeed {\it essential} in producing large-scale magnetic fields of
the general type normally observed. If we artificially take
$\alpha=0$, the field rapidly winds up. In contrast, we
have demonstrated the organizational effects of the large-scale dynamo
action presented in this paper.

A noteworthy feature of the models is the existence of long-lived
large-scale field reversals for some choices of parameters. This may
explain the possible existence of such a reversal in the Milky
Way. It is too early to judge whether such features are realistic --
existing observational data of required quality may be too sparse to
assess this.

The accuracy of determination of the dynamo-governing parameters
from observational data is substantially lower than the variation
required to switch between models with and without reversals. We
conclude that dynamo theory is presently unable to predict if and
how many reversals are expected in a given galaxy, due to
insufficient information about galactic hydrodynamics and inherent
uncertainty in both the initial magnetic fields and those resulting
from local small-scale dynamo action.

We have constructed synthetic maps of polarized radio synchrotron
emission and Faraday rotation measures for various evolutionary
epochs. The maps for the present-day epoch are similar to those
observed in many nearby spiral galaxies. Spiral arms, bars and
outflows into the halo are known to affect magnetic field patterns
(e.g. Beck 2010), but these phenomena are not included in our models
which assume an axisymmetric gas flow in the galaxy plane.

Synchrotron emission from distant galaxies show that
total magnetic fields exist in distant galaxies (Murphy
2009), but the sensitivity of present-day radio telescopes does not
allow measurement of the large-scale field and its
patterns. The SKA and its precursor telescopes offer the
opportunity to measure field patterns for a large number of galaxies
at various evolutionary stages. Significant polarized
emission, the signature of ordered magnetic fields, is expected at
redshifts $\le4$ and Faraday rotation measures (RM), the
signature of large-scale regular fields, at redshifts
{$\le3$}. Large-scale field reversals are more likely to
occur in galaxies with large dynamo number, e.g. galaxies with rapid
rotation. Failure to detect large-scale RM patterns hints at a major
encounter or merger with another galaxy, which would destroy much of
the ordered structure from earlier evolution, and so reset the clock
for the evolution of the large-scale field.

\acknowledgements TGA acknowledges support by the DFG--SPP project under
grant 566960. This work is supported by the European Community
Framework Programme 6, Square Kilometre Array Design Study (SKADS),
and the DFG--RFBR project under grant 08-02-92881. The Royal Society
supported a visit to Russia by DM. RS acknowledges the grant YD-4471.2011.1 by the Council
of the President of the Russian Federation and the RFBR grant
11-01-96031-ural.

%\newpage
%%%%%%%%%%%%%%%%%%%%%%%%%%%%%%%%%%%%%%%%%%%%%%%%%%%%%%

\end{document}